\documentclass[%
 reprint,
runinaddress,
 amsmath,amssymb,
 aps,
prc,
]{revtex4-1}

\usepackage{graphicx}
\usepackage{dcolumn}
\usepackage{bm}
\usepackage{lineno}

\usepackage{booktabs}
\newcolumntype{C}[1]{>{\centering\arraybackslash}p{#1}}
\AtBeginDocument{
\heavyrulewidth=.08em
\lightrulewidth=.05em
\cmidrulewidth=.03em
\belowrulesep=.65ex
\belowbottomsep=0pt
\aboverulesep=.4ex
\abovetopsep=0pt
\cmidrulesep=\doublerulesep
\cmidrulekern=.5em
\defaultaddspace=.5em
}

\begin{document}

\preprint{APS/123-QED}

\title{Total absorption $\gamma$-ray spectroscopy of the $\beta$ decays of $^{96\text{gs,m}}$Y}

\author{V. Guadilla}
\altaffiliation{%
Corresponding author: vguadilla@fuw.edu.pl 
\newline
 Present address: Faculty of Physics, University of Warsaw, 02-093 Warsaw, Poland
}
\affiliation{%
 Subatech, IMT-Atlantique, Universit\'e de Nantes, CNRS-IN2P3, F-44307, Nantes, France
}
\author{L. Le Meur}  
\affiliation{%
 Subatech, IMT-Atlantique, Universit\'e de Nantes, CNRS-IN2P3, F-44307, Nantes, France
}
\author{M. Fallot}  
\affiliation{%
 Subatech, IMT-Atlantique, Universit\'e de Nantes, CNRS-IN2P3, F-44307, Nantes, France
}
\author{A. Algora}%
\altaffiliation{%
 Institute of Nuclear Research of the Hungarian Academy of Sciences, Debrecen H-4026, Hungary
}
\affiliation{%
 Instituto de F\'isica Corpuscular, CSIC-Universidad de Valencia, E-46071, Valencia, Spain
}
\author{J. L. Tain}%
\affiliation{%
 Instituto de F\'isica Corpuscular, CSIC-Universidad de Valencia, E-46071, Valencia, Spain
}
\author{J. Agramunt}  
\affiliation{%
 Instituto de F\'isica Corpuscular, CSIC-Universidad de Valencia, E-46071, Valencia, Spain
}
\author{J. \"Ayst\"o}  
\affiliation{%
 University of Jyv\"askyl\"a, FI-40014, Jyv\"askyl\"a, Finland
}
\author{J. A. Briz}  
\affiliation{%
 Subatech, IMT-Atlantique, Universit\'e de Nantes, CNRS-IN2P3, F-44307, Nantes, France
}
\author{T. Eronen}  
\affiliation{%
 University of Jyv\"askyl\"a, FI-40014, Jyv\"askyl\"a, Finland
}
\author{M. Estienne}  
\affiliation{%
 Subatech, IMT-Atlantique, Universit\'e de Nantes, CNRS-IN2P3, F-44307, Nantes, France
}
\author{L. M. Fraile}  
\affiliation{%
Grupo de F\'isica Nuclear $\&$ IPARCOS, Universidad Complutense de Madrid, CEI Moncloa, E-28040, Madrid, Spain
}
\author{E. Ganio\u{g}lu}  
\affiliation{%
Department of Physics, Istanbul University, 34134, Istanbul, Turkey
}
\author{W. Gelletly}  
\affiliation{%
Department of Physics, University of Surrey, GU2 7XH, Guildford, UK
} 
\author{\\ L. Giot}  
\affiliation{%
 Subatech, IMT-Atlantique, Universit\'e de Nantes, CNRS-IN2P3, F-44307, Nantes, France
}
\author{D. Gorelov}  
\author{J. Hakala} 
\author{A. Jokinen}
\affiliation{%
 University of Jyv\"askyl\"a, FI-40014, Jyv\"askyl\"a, Finland
}\author{D. Jordan} 
\affiliation{%
 Instituto de F\'isica Corpuscular, CSIC-Universidad de Valencia, E-46071, Valencia, Spain
}  
\author{A. Kankainen}  
\author{V. Kolhinen}  
\affiliation{%
 University of Jyv\"askyl\"a, FI-40014, Jyv\"askyl\"a, Finland
}
\author{J. Koponen}  
\affiliation{%
 University of Jyv\"askyl\"a, FI-40014, Jyv\"askyl\"a, Finland
}
\author{M. Lebois}  
\affiliation{%
Institut de Physique Nucl\`eaire d'Orsay, 91406, Orsay, France
}
\author{T. Martinez}  
\affiliation{%
Centro de Investigaciones Energ\'eticas Medioambientales y Tecnol\'ogicas, E-28040, Madrid, Spain
}
\author{M. Monserrate}  
\author{A. Montaner-Piz\'a}  
\affiliation{%
 Instituto de F\'isica Corpuscular, CSIC-Universidad de Valencia, E-46071, Valencia, Spain
}
\author{I. Moore}  
\affiliation{%
 University of Jyv\"askyl\"a, FI-40014, Jyv\"askyl\"a, Finland
}
\author{E. N\'acher}  
\affiliation{%
 Instituto de F\'isica Corpuscular, CSIC-Universidad de Valencia, E-46071, Valencia, Spain
}
\author{S. E. A. Orrigo}  
\affiliation{%
 Instituto de F\'isica Corpuscular, CSIC-Universidad de Valencia, E-46071, Valencia, Spain
}
\author{H. Penttil\"a}  
\affiliation{%
 University of Jyv\"askyl\"a, FI-40014, Jyv\"askyl\"a, Finland
}
\author{I. Pohjalainen}  
\affiliation{%
 University of Jyv\"askyl\"a, FI-40014, Jyv\"askyl\"a, Finland
}
\author{A. Porta}  
\affiliation{%
 Subatech, IMT-Atlantique, Universit\'e de Nantes, CNRS-IN2P3, F-44307, Nantes, France
}
\author{J. Reinikainen}  
\author{M. Reponen}  
\author{S. Rinta-Antila}  
\affiliation{%
 University of Jyv\"askyl\"a, FI-40014, Jyv\"askyl\"a, Finland
}
\author{B. Rubio}  
\affiliation{%
 Instituto de F\'isica Corpuscular, CSIC-Universidad de Valencia, E-46071, Valencia, Spain
}
\author{K. Rytk\"onen}  
\affiliation{%
 University of Jyv\"askyl\"a, FI-40014, Jyv\"askyl\"a, Finland
}
\author{A. Cucoanes}  
\affiliation{%
 Subatech, IMT-Atlantique, Universit\'e de Nantes, CNRS-IN2P3, F-44307, Nantes, France
}
\author{T. Shiba}  
\affiliation{%
 Subatech, IMT-Atlantique, Universit\'e de Nantes, CNRS-IN2P3, F-44307, Nantes, France
}
\author{V. Sonnenschein}  
\affiliation{%
 University of Jyv\"askyl\"a, FI-40014, Jyv\"askyl\"a, Finland
}
\author{A. A. Sonzogni}  
\affiliation{%
NNDC, Brookhaven National Laboratory, Upton, NY 11973-5000, USA
}
\author{E. Valencia}  
\affiliation{%
 Instituto de F\'isica Corpuscular, CSIC-Universidad de Valencia, E-46071, Valencia, Spain
}
\author{V. Vedia}  
\affiliation{%
Grupo de F\'isica Nuclear $\&$ IPARCOS, Universidad Complutense de Madrid, CEI Moncloa, E-28040, Madrid, Spain
}
\author{A. Voss} 
\affiliation{%
 University of Jyv\"askyl\"a, FI-40014, Jyv\"askyl\"a, Finland
}
\author{J. N. Wilson}
\affiliation{%
Institut de Physique Nucl\`eaire d'Orsay, 91406, Orsay, France
}
\author{A. -A. Zakari-Issoufou} 
\affiliation{%
 Subatech, IMT-Atlantique, Universit\'e de Nantes, CNRS-IN2P3, F-44307, Nantes, France
}

\date{\today}

\begin{abstract}
 
The $\beta$ decays of the ground state (gs) and isomeric state (m) of $^{96}$Y have been studied with the total absorption $\gamma$-ray spectroscopy technique at the Ion Guide Isotope Separator On-Line facility. The separation of the 8$^{+}$ isomeric state from the 0$^{-}$ ground state was achieved thanks to the purification capabilities of the JYFLTRAP double Penning trap system. The $\beta$-intensity distributions of both decays have been independently determined. In the analyses the de-excitation of the 1581.6~keV level in $^{96}$Zr, in which conversion electron emission competes with pair production, has been carefully considered and found to have significant impact on the $\beta$-detector efficiency, influencing the $\beta$-intensity distribution obtained. Our results for $^{96\text{gs}}$Y (0$^+$) confirm the large ground state to ground state $\beta$-intensity probability, although a slightly larger value than reported in previous studies was obtained, amounting to $96.6_{-2.1}^{+0.3}\%$ of the total $\beta$ intensity. Given that the decay of $^{96\text{gs}}$Y is the second most important contributor to the reactor antineutrino spectrum between 5 and 7~MeV, the impact of the present results on reactor antineutrino summation calculations has been evaluated. In the decay of $^{96\text{m}}$Y (8$^{+}$), previously undetected $\beta$ intensity in transitions to states above 6~MeV has been observed. This shows the importance of total absorption $\gamma$-ray spectroscopy measurements of $\beta$ decays with highly fragmented de-excitation patterns. $^{96\text{m}}$Y (8$^{+}$) is a major contributor to reactor decay heat in uranium-plutonium and thorium-uranium fuels around 10~s after fission pulses, and the newly measured average $\beta$ and $\gamma$ energies differ significantly from the previous values in evaluated databases. The discrepancy is far above the previously quoted uncertainties. Finally, we also report on the successful implementation of an innovative total absorption $\gamma$-ray spectroscopy analysis of the module-multiplicity gated spectra, as a first proof of principle to distinguish between decaying states with very different spin-parity values.

\end{abstract}

\maketitle

\section{Introduction}

The $\beta$ decay of neutron-rich nuclei produced in the fission of nuclear reactor fuel is the source of a large part of the reactor decay heat, producing antineutrinos and $\beta$-delayed neutrons. The accurate prediction of the reactor decay heat is crucial for safe control of nuclear reactors, as it is the dominant source of energy when reactors are turned off, as well as for reactor waste management. A summation method based on nuclear data can be employed to compute the decay heat, offering a flexible approach that allows one to make predictions for new reactors and new fuel compositions. The understanding of the reactor antineutrino spectrum is important for reactor-based experiments on fundamental neutrino physics~\cite{DoubleChooz,DayaBay,Reno} and for reactor monitoring in safeguard inspection~\cite{non_proliferation_Nature}. Regarding antineutrinos, two computing approaches are used:
a) the summation method~\cite{neutrinos_PRL,Sonzogni_summation} and b) the conversion of integral $\beta$-spectrum measurements for the main fissile isotopes~\cite{Mueller_neutrinos,Huber_conversion2}. Recently the observation of discrepancies between the experimental spectra and the calculated ones in terms of the absolute flux (called the ``reactor antineutrino anomaly'')~\cite{Anomaly} and in spectral shape (called the ``shape anomaly'')~\cite{RENO_shoulder,DayaBay_shoulder,DoubleChooz_shoulder} has triggered new efforts to improve the accuracy of antineutrino spectrum calculations using the summation method.

The summation methods applied to reactor decay heat and reactor antineutrino spectrum calculations depend on the quality of the available nuclear data. One of the ingredients used as input are the $\beta$-intensity probabilities of populating the daughter levels in the $\beta$ decay of each fission fragment. In particular, the average $\gamma$ and $\beta$ energies, calculated from the $\beta$-decay probabilities, are employed to determine the decay heat as a function of time by summing the energy released by the decay of each nucleus weighted by its corresponding activity at that time. In the same way, the antineutrino spectrum associated with the decay of each fission product is determined by using the $\beta$-intensity probabilities. These individual spectra are summed, assuming $\beta$-transition types and weighted by their corresponding activity, to calculate the total reactor antineutrino spectrum at a given time. 

It is important to note that many of the fission fragments of interest for reactor calculations are short-lived nuclei and have large $\beta$-decay energy windows $Q_{\beta}$. Such cases are associated with more complex decay patterns, involving high $\gamma$-multiplicity cascades in the de-excitation of the numerous levels fed in $\beta$ decay. Traditional high-resolution $\gamma$-spectroscopy approaches based on HPGe detectors are known to be impaired by their limited efficiency. This leads to the non-detection of part of these $\gamma$ cascades, with the resulting underestimation of the $\beta$ feeding of levels at high excitation energy, called the Pandemonium effect~\cite{Pandemonium}. The Total Absorption $\gamma$-ray Spectroscopy (TAGS) technique offers an alternative high-efficiency approach to overcome this problem~\cite{Rubio_tas}, by covering almost the full solid angle with large scintillator crystals. In recent years, TAGS has proven to be a suitable tool to investigate the $\beta$ decay of neutron-rich nuclei, with enough sensitivity even to detect $\gamma$ rays from neutron-unbound states~\cite{vTAS_PRL,vTAS_PRC,SUN_neutron_PRL,PRC_BDN}.

The summation method allows one to identify the nuclei that contribute most to the reactor antineutrino spectrum and to the reactor decay heat~\cite{neutrinos_PRL,Sonzogni_summation,Zak_PRL}. This was first done in the case of reactor decay heat in the pioneering work of Yoshida and Nichols and the Working Party on International Evaluation Co-operation of the NEA Nuclear Science Committee (WPEC 25) group~\cite{NEA_IAEA_DecayHeat1}, highlighting the important role of the Pandemonium effect in the decay data of important contributors and providing lists of nuclei that would deserve new TAGS measurements. The first TAGS measurements of some of these key nuclei had a very large impact on the decay heat after a thermal fission pulse of $^{239}$Pu, thus solving a long-lasting discrepancy between integral measurements and summation calculations~\cite{DecayHeat}. It was evidenced in~\cite{neutrinos_PRL} that the reactor antineutrino spectra computed with the summation method also suffer from the Pandemonium decay data, similarly to the decay heat. Since then, many other studies~\cite{Zak_PRL, vTAS_PRC, MTAS_neutrino_PRL, MTAS_neutrino_DH_PRL, PRL_Nb, IAEA2015} have also helped to improve reactor summation calculations. 

Recently, the Nantes summation method for the calculation of the reactor antineutrino spectrum was updated with the published results of the TAGS campaigns carried out by our collaboration during the last decade, significantly improving the agreement between the experimental reactor antineutrino results and the calculation of the absolute flux ~\cite{PRL_Magali} and showing that the remaining flux discrepancy should be even further reduced with additional Pandemonium-free data. This reinforces the need for decay data free from the Pandemonium effect to improve the summation method. For a review of the impact of our TAGS campaigns during the last decade, we refer the interested  reader to the recent review article~\cite{TAGS_review}.

In this article we present the study of the $\beta$ decays of the ground state (gs) and isomeric state (m) of $^{96}$Y. These decays are estimated to produce almost 5\% of the decay heat around 10~s after fission in $^{235}$U~\cite{NEA_IAEA_DecayHeat1}. The TAGS study of the decay of the 0$^-$ ground state was ranked as priority two for U/Pu and Th/U fuels by a committee of experts of the International Atomic Energy Agency (IAEA)~\cite{IAEA2015}, while the TAGS study of the decay of the 8$^+$ isomeric state was considered priority one for Th/U fuel. The TAGS measurements of these cases are thus of enormous importance in order to increase confidence in decay heat calculations. In addition, the decay of the 0$^-$ ground state is one of the main contributors to the reactor antineutrino spectrum in the region of the spectral shape distortion between 5 and 7~MeV, adding almost 11$\%$ of the antineutrino spectrum of a pressurized water reactor (PWR) in the 5-6~MeV energy range and 14$\%$ between 6 and 7~MeV~\cite{Zak_PRL}. Its TAGS measurement was ranked of the highest priority by the IAEA~\cite{IAEA2015} for the improvement of the reactor antineutrino spectrum. 

On the other hand, the decay of the 8$^+$ isomeric state is also important if we are to understand the structure of the daughter nucleus, $^{96}$Zr, which lies in a region of shape transition~\cite{ShapeZr2016} and emergence of shape-coexisting states and intertwined quantum phase transitions~\cite{Zr_2021_theo}. In the $\beta$ decay of $^{96\text{m}}$Y moderately high-spin levels are expected to be accessed, in contrast with the decay of the ground state that mainly populates low-spin levels. 
Shape coexistence in $^{96}$Zr has been established in experiment recently~\cite{Shape96Zr} triggering many theoretical works to elucidate the properties of this nucleus and reinforcing the interest in even-even Zr isotopes~\cite{PhysRevC_Zr_2019,Zr_Gavrielov_2019,PhysRevC_Zr_2020}. In particular, recent beyond-mean field calculations~\cite{Petro96Y_2020} have studied the triple shape coexistence of the 0$^{+}$ states in $^{96}$Zr, as well as the dominance of a prolate configuration in the 8$^+$ state in $^{96}$Zr, predominately populated in the decay of $^{96\text{m}}$Y.

In recent previous measurements of the $\beta$ decay of $^{96}$Y either the 8$^+$ isomer was not produced and could not be studied~\cite{MTAS_neutrino_PRL}, or the $\beta$ decays of the ground state and the isomer were mixed~\cite{PLB_96Y_2021}. Therefore, the analysis of their decay patterns had to rely to some extent on the previous high-resolution spectroscopy measurements~\cite{Mach_96Y} to disentangle the two components, especially at high energy. The measurements of the $\beta$ decays of the ground state and the isomer of $^{96}$Y presented here are unambiguously separated thanks to the use of the JYFLTRAP double Penning trap system~\cite{JYFLTRAP}, as will be described later. We have also carefully taken into account in the TAGS analyses the electric monopole ($E0$) transition from the 1581.6~keV state in $^{96}$Zr and we show the impact on the obtained results.

The present article is organized as follows: in Sec.~\ref{Exp} we give details about the experimental TAGS measurements and the analysis procedure is detailed in Sec.~\ref{TAGS}, where the main results are discussed. In particular, the first application of a novel analysis approach for module-multiplicity gated spectra is presented in Sec.~\ref{Mm_ana}, the impact on reactor summation calculations is discussed in Sec.~\ref{Reactor} and in Sec.~\ref{beta_spec} we determine the $\beta$-energy spectrum of the decay of $^{96\text{gs}}$Y from the results of this work and we compare it with previous $\beta$-spectra measurements. Finally, general conclusions are drawn in Sec.~\ref{Conclusions}.

\section{Experiment}\label{Exp}

\begin{figure}[h]
\begin{center}
\includegraphics[width=0.5 \textwidth]{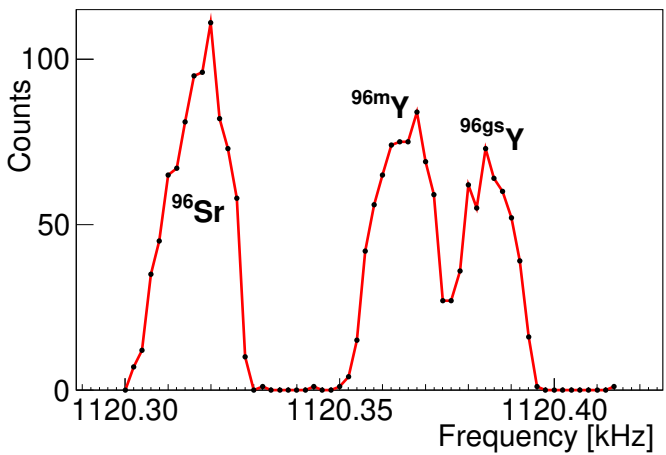} 
\caption{JYFLTRAP purification trap mass scans for A = 96. The frequency is selected to extract the isobar of interest from the trap.}
\label{96Y_mass}
\end{center}
\end{figure}

The measurement of the decays of $^{96\text{gs,m}}$Y was performed as part of a TAGS campaign at the upgraded IGISOL IV facility of the University of Jyv\"askyl\"a (Finland) \citep{Moore_IGISOLIV}. The segmented Decay Total Absorption $\gamma$-ray Spectrometer (DTAS) \citep{DTAS_design}, composed of eighteen NaI(Tl) crystals, was employed in coincidence with a thin plastic $\beta$ detector (see~\citep{NIMB_DTAS} for more details about the experimental setup). Proton-induced fission on natural uranium produced the nuclear species of interest, extracted with the fission ion guide technique, separated in mass with the mass separator magnet and further purified in the JYFLTRAP double Penning trap system~\cite{JYFLTRAP} to select the isobaric component of interest. The half-lives of the two $\beta$-decaying states, $^{96\text{gs,m}}$Y, are 5.34(5) and 9.6(2)~s~\cite{NDS_96}, respectively, and the $Q_{\beta}$ value of the decay of the ground state is 7103(6)~keV~\cite{Qval_NNDC}. In this case no $\beta$-delayed neutron branch is energetically possible. For the energy of the 8$^{+}$ isomer in this work we use the 1540(9)~keV value from NUBASE 2020~\cite{NUBASE2020} (also quoted in~\cite{NUBASE2012,NUBASE2016}) based on a precision mass measurement with JYFLTRAP~\cite{96Y_mass_2007_IGISOL}. A 1540.5(4)~keV value has recently been deduced from the study of a (6$^{+}$) 181(9)~ns isomer at 1655~keV excitation energy in $^{96}$Y~\cite{PRC_2020_Krakow}, thus reinforcing the current NUBASE value. Note that in NUBASE 2003~\cite{NUBASE2003} the quoted value for the 8$^{+}$ isomer was 1140(30)~keV, which is the value available in the Evaluated Nuclear Structure Data File (ENSDF)~\cite{NDS_96} and National Nuclear Data Center (NNDC)~\cite{NNDC} databases. In our experiment, the energy difference between the ground state and the 8$^{+}$ isomeric state of $^{96}$Y was resolved with JYFLTRAP using the buffer-gas cooling technique~\cite{Savard1991} in the first trap, known as the purification trap. An example of such a mass scan is shown in Fig.~\ref{96Y_mass}. 

\begin{figure}[h]
\begin{center}
\includegraphics[width=0.5 \textwidth]{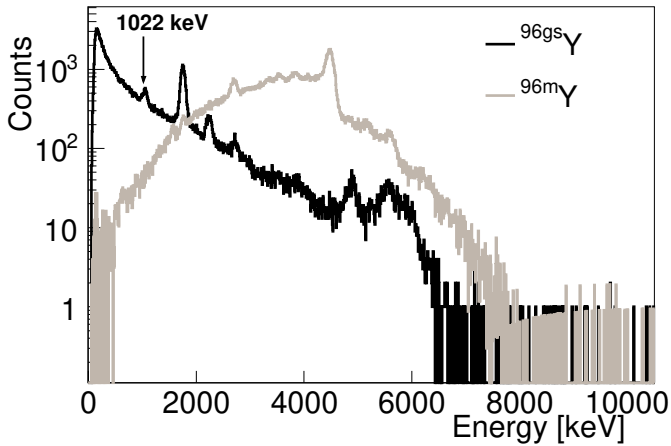} 
\caption{Experimental $\beta$-gated spectra of the measurements for $^{96\text{gs}}$Y (black) and $^{96\text{m}}$Y (grey) free of contaminants. The sum peak of the two 511~keV $\gamma$ rays emitted in the positron annihilation from the pair emission of the 1581.6~keV level in $^{96}$Zr is indicated (see text for details).}
\label{96Y_spectra}
\end{center}
\end{figure}

The $\beta$-gated total absorption spectrum of the DTAS detector was reconstructed offline following the procedure described in Ref.~\cite{NIMA_DTAS}, by summing the signals from the individual modules in coincidence with the $\beta$-plastic detector. The experimental $\beta$-gated spectra corresponding to the decays of $^{96\text{gs,m}}$Y are presented in Fig.~\ref{96Y_spectra} free of contaminants. The contaminants which were subtracted from the spectra include several contributions. In both cases the summing-pileup distortion was considered, and it was calculated and normalized as explained in Ref.~\cite{NIMA_DTAS}, based on the Monte Carlo (MC) method developed by the group of Valencia~\cite{TAS_pileup}. In the decay of the ground state this was the only source of contamination subtracted. In the decay of the 8$^{+}$ isomeric state, in addition to the summing-pileup contribution, we also considered a contamination coming from the decays of $^{96\text{gs}}$Y and $^{96}$Sr, due to problems in the purification in JYFLTRAP for some experimental runs. The decay of $^{96}$Sr was measured in the same experimental campaign and it was subtracted from the 8$^{+}$ isomeric spectrum  by normalizing with the peak at 931.7~keV, coming from the de-excitation of the most populated level in the decay of $^{96}$Sr. The contamination of $^{96\text{gs}}$Y in the $^{96\text{m}}$Y spectrum was normalized by matching the low energy region of the spectrum, associated with the penetration in DTAS of high-energy $\beta$ particles from the ground state to ground state transition (only allowed in the decay of $^{96\text{gs}}$Y).

\section{Analysis and results}\label{TAGS}

For the TAGS analyses of the experimental spectra, we followed the method developed by the group of Valencia~\cite{TAS_MC,TAS_algorithms,TAS_decaygen} to determine the $\beta$ intensities. The inverse problem $d_i= \sum_j^{levels} R_{ij}(B) f_j +C_i$ has to be solved in order to determine $f_j$, the number of decay events that feed level $j$ in the daughter nucleus, where $d_i$ represents the number of counts in channel $i$ of the total absorption spectrum, $C_i$ represents all contaminants in channel $i$, and $R_{ij}$ is the response function of the total absorption spectrometer. The response function depends on the branching ratios ($B$) for the different de-excitation paths of the states populated in the decay. At low excitation energies, these branching ratios are taken from the literature, assuming a good knowledge of the level scheme. When the information starts to be incomplete, we introduce a continuum region of 40~keV bins, where the branching ratios are calculated based on a statistical model~\cite{TAS_decaygen}.

With regard to the decays of $^{96\text{gs,m}}$Y into $^{96}$Zr, we have considered the known level scheme in the daughter nucleus up to 4389.5~keV excitation energy (an 8$^+$ level strongly fed in the decay of the 8$^+$ $^{96\text{m}}$Y isomer~\cite{Julich_96mY}), taking the ENSDF data~\cite{NDS_96} as input. For those levels without firm spin-parity assignment in the ENSDF database, we have chosen the values according to the recommendations given by the Reference Input Parameter Library (RIPL-3)~\cite{RIPL-3} up to 3772.2~keV excitation energy, except for the level at 3309.19~keV (a 6$^+$ is recommended by RIPL), because the analysis of the decay of the 8$^{+}$ isomer was improved with a 4$^+$ value. From 3772.2~keV excitation energy up to 4389.5~keV RIPL does not give any recommendation for those levels without spin-parity value assigned and our choices among the possible values quoted in the ENSDF evaluation are justified as follows: the $1^+$ selection for the levels at 3947.19 and 4037.89~keV as well as the $1^-$ value for the level at 4132.4~keV improve the reproduction of the experimental spectra for the decay of the 0$^-$ $^{96\text{gs}}$Y. The level at 4261.3~keV excitation energy has been chosen to be a 6$^+$ although the 5$^+$ value was also considered for the calculation of the uncertainties (with no effect in any of the analyses). There are three levels without any tentative spin-parity value. Those at 3924.6 and 4024.5~keV have been chosen not to be directly fed in any of the decays (assuming 4$^+$ and 5$^+$ values, respectively), since we have verified that those levels are not seen in the spectra. Finally, the level lying at 3865.16~keV excitation energy has been chosen to be a 2$^+$, because the reproduction of the experimental $^{96\text{gs}}$Y spectrum was found to improve slightly when this level is directly fed (first forbidden transition). The possible influence of the spin-parity values of these three levels has been investigated and in the evaluation of the uncertainties alternative 3$^{\pm}$, 4$^{\pm}$ and 5$^{\pm}$ values were also considered for them. It turned out that any effects are negligible as we shall see later.

From 4389.5~keV excitation energy up to the $Q_{\beta}$ value, the branching ratios have been calculated based on a statistical model that uses the parameters given in Table~\ref{parameters_PSF} as input, and taken from RIPL-3~\cite{RIPL-3}. As presented in Table~\ref{parameters_PSF}, two $\beta$-deformation parameters used for the Photon Strength Function (PSF) calculation have been considered: the reference one based on experimental results~\cite{DeformationPar_exp} and an alternative value coming from the finite range droplet model (FRDM) calculations available at RIPL-3, which predict a larger deformation. The level density parameter ``a'' at the neutron binding energy employed for the calculation of the E1 $\gamma$-strength function is the one obtained with the TALYS code~\cite{TALYS}, since no value for $^{96}$Zr is available at RIPL-3. We have used the nuclear level density from the Hartree-Fock-Bogoliubov (HFB) plus combinatorial model~\cite{Gorieli1,Gorieli2}, retrieved from RIPL-3.

\begin{table*}[!t]
\begin{ruledtabular}
\begin{minipage}{\textwidth}
 \begin{center}
  \begin{tabular}{@{}C{2cm}C{2cm}C{1cm}C{1cm}C{1.5cm}C{1cm}C{1cm}C{1cm}C{1cm}C{1cm}C{1cm}@{}}
  \\ \cmidrule(lr){1-11}
    Level-density parameter
    & Deformation parameter
    & \multicolumn{9}{c}{Photon strength function parameters}
    \\ \cmidrule(lr){3-11}
    &  & \multicolumn{3}{c}{E1} & \multicolumn{3}{c}{M1} & \multicolumn{3}{c}{E2}   \\
    \cmidrule(r){1-1}\cmidrule(lr){2-2}\cmidrule(lr){3-5}\cmidrule(lr){6-8}\cmidrule(l){9-11}
    a & $\beta$  & E & $\Gamma$ & $\sigma$ & E & $\Gamma$ & $\sigma$ & E & $\Gamma$ & $\sigma$  \\
    $[$MeV$^{-1}$] &  & [MeV] & [MeV] & [mb] & [MeV] & [MeV] & [mb] & [MeV] & [MeV] & [mb]  \\
    \\ \cmidrule(lr){1-11}
    &  & 15.720 & 4.878 & 63.953 & & & & & \\[-1ex]
    & \raisebox{1.5ex}{0.08} & 16.832 & 5.554 & 112.329 & & & \raisebox{1.5ex}{0.681} & & & \\
    
    \cmidrule(lr){2-5}  \cmidrule(lr){8-8} \\ [-5ex]
 14.57370 & & & & & 8.968 & 4.000 & &  13.780 & 4.958 & 2.010 \\ 
    
    &  & 14.655 & 4.269 & 73.064 & & & & & \\[-1ex]
    & \raisebox{1.5ex}{0.217} & 17.598 & 6.044 & 103.216 & & & \raisebox{1.5ex}{0.736} & & & \\
  \end{tabular}
  \caption{\label{parameters_PSF} Parameters used in the statistical model calculation of the branching ratio matrix (B) of the daughter nucleus $^{96}$Zr. The upper set of $E1$ parameters is calculated with an experimental deformation parameter, whereas the lower one is calculated with a deformation parameter coming from FRDM calculations.}
 \end{center}
\end{minipage}
\end{ruledtabular}
\end{table*}

Once the branching ratio matrix for each decay was determined, the response function was calculated by means of MC simulations~\cite{TAS_MC}, using the Geant4 simulation package \cite{GEANT4}. For this, monoenergetic $\gamma$-ray MC responses are normally folded with the response to the $\beta$ continuum for each level~\cite{TAS_MC}. In the next section we will comment on a slight modification of this procedure introduced to treat properly the de-excitation of the 1581.6~keV level. The MC simulations were validated by comparison with measurements of well-known radioactive sources ($^{60}$Co, $^{137}$Cs, $^{22}$Na, $^{24}$Na, and a mixture of $^{152}$Eu and $^{133}$Ba)~\cite{NIMA_DTAS}. Finally, the $\beta$-intensity distributions were determined by applying an expectation maximization (EM) algorithm~\cite{TAS_algorithms}.

\subsection{De-excitation of the 1581.6~keV level}\label{1581}

\begin{figure}[h]
\begin{center}
\includegraphics[width=0.5 \textwidth]{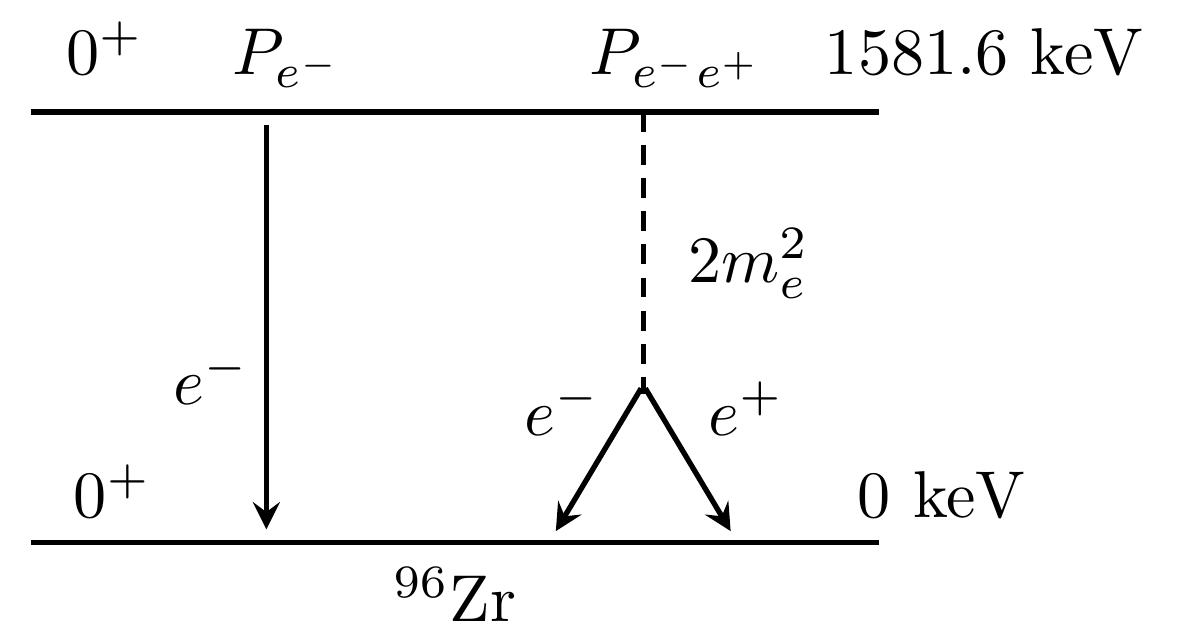} 
\caption{Scheme of the possible $E0$ de-excitations of the level at 1581.6~keV excitation energy in $^{96}$Zr. Pair production occurs with a probability $P_{e^-e^+}$.}
\label{96gsY_scheme}
\end{center}
\end{figure}

In the calculation of the response matrices of the decays of $^{96\text{gs,m}}$Y one special feature of the level scheme of $^{96}$Zr was taken into account. The de-excitation of the 0$^+$ level at 1581.6~keV, presented in Fig.~\ref{96gsY_scheme}, occurs by means of conversion electrons in competition with pair production, due to the fact that the energy involved exceeds the pair production threshold 2$m_{e}$ (1022~keV). A careful study of the absolute $E0$ intensity was performed by Mach \textit{et al.}~\cite{Mach_96Y}, where a probability for pair production $P_{e^-e^+}$=0.170 was used, based on the Wilkinson formulation~\cite{Wilkinson_NPA, Wilkinson_NIM}. In the present work, we took $P_{e^-e^+}$=0.143 from the BrIcc (v2.3) conversion coefficient calculator~\cite{BRICC_NIMA,BRICC}, compatible with the value used by Mach \textit{et al.}~\cite{Mach_96Y} within an uncertainty of 25\%. A recent tabulation for the upcoming BrIcc (v3.1) gives a value for $P_{e^-e^+}$ of 0.138~\cite{BRICC_2020}. No experimental value is available for $P_{e^-e^+}$ in this case, but in the case of $^{90}$Zr, with an $E0$ transition of 1760.7~keV, BrIcc (v2.3) gives a probability for pair emission with a maximum 15\% discrepancy with respect to experimental values (considering only $K$ shell electrons).

\begin{figure}[h]
\begin{center}
\includegraphics[width=0.5 \textwidth]{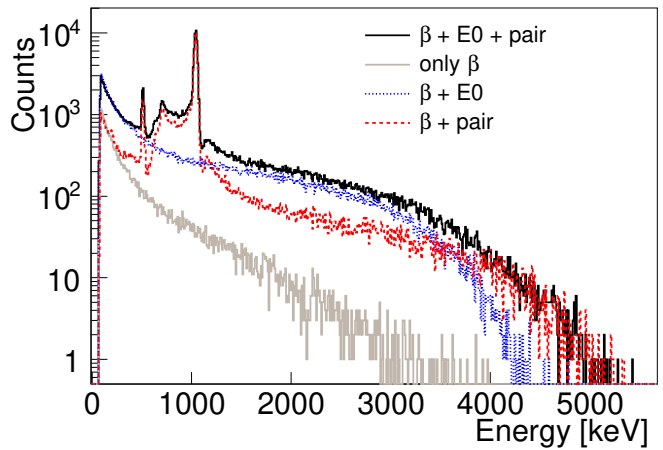} 
\caption{MC response for the 1581.6~keV level of the $\beta$-gated TAGS spectrum. 
The solid grey line shows the normal penetration of $\beta$ particles in the spectrometer, whereas the solid black line shows the effect of also taking into account conversion electrons and the pair production for the de-excitation of this level. For more detail we add the dotted blue line that corresponds to considering only conversion electrons together with $\beta$ particles, and the red dashed line showing the contribution of pair production and $\beta$ electrons.}
\label{1582_response}
\end{center}
\end{figure}

To take into account the response of this level when constructing the response function, we have replaced the usual $\gamma$ response by a MC simulation of this de-excitation pattern. In this MC simulation we generated an electron-positron pair with a probability $P_{e^-e^+}$ and a conversion electron with a probability 1-$P_{e^-e^+}$. The simulated energy of the conversion electron corresponded to the energy of the level minus the binding energy of the $K$ electron in Zr (18.0~keV). The energy of the pair corresponded to the energy of the level minus 2$m_e$ and it was randomly shared between the electron and positron, that were simulated as being emitted back to back. For the case of direct $\beta$ feeding the resulting MC response for this level is shown in Fig.~\ref{1582_response}, in comparison with the response that only takes into account the $\beta$ particles emitted.

\begin{figure}[h]
\begin{center}
\includegraphics[width=0.5 \textwidth]{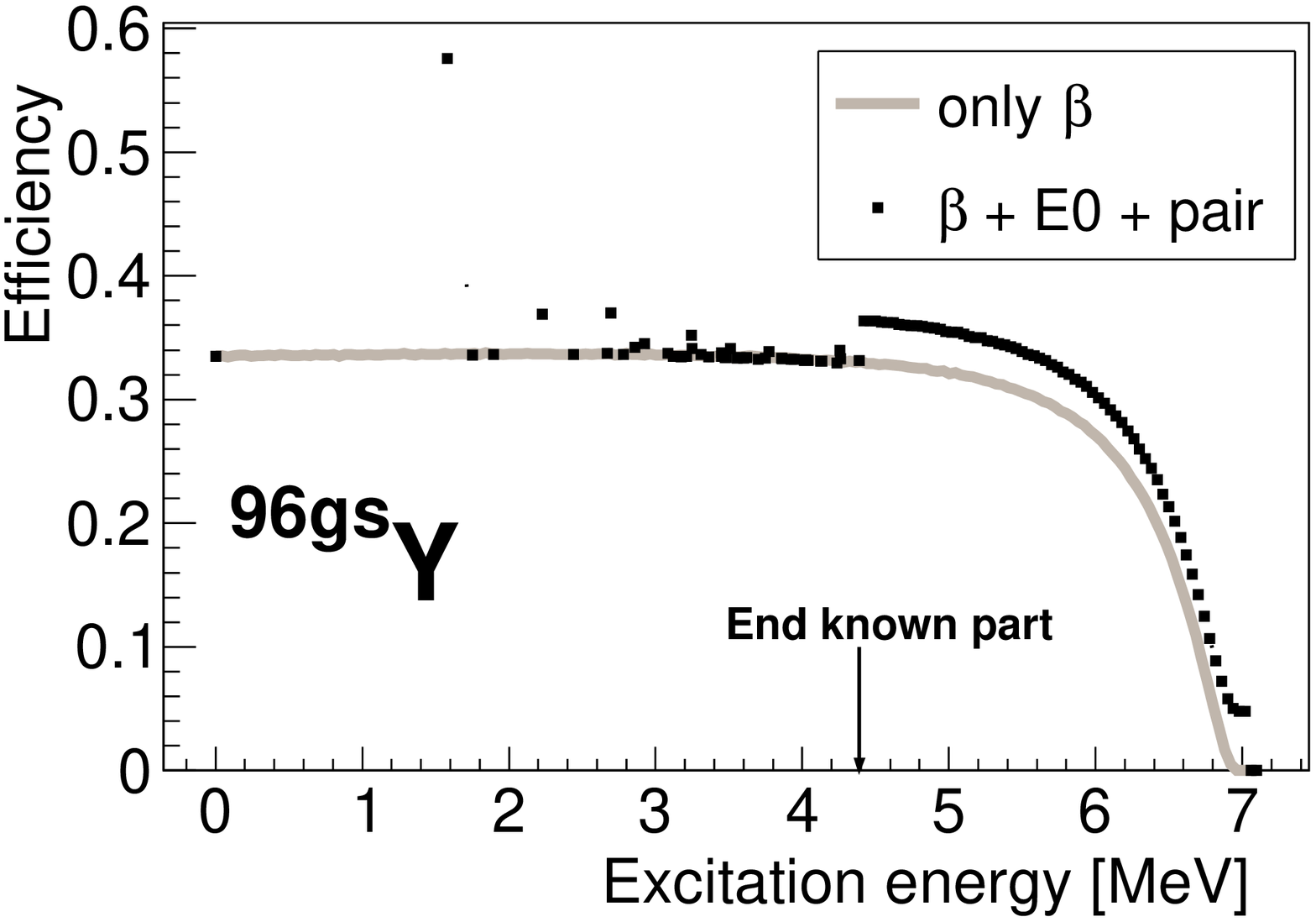} \\

\includegraphics[width=0.5 \textwidth]{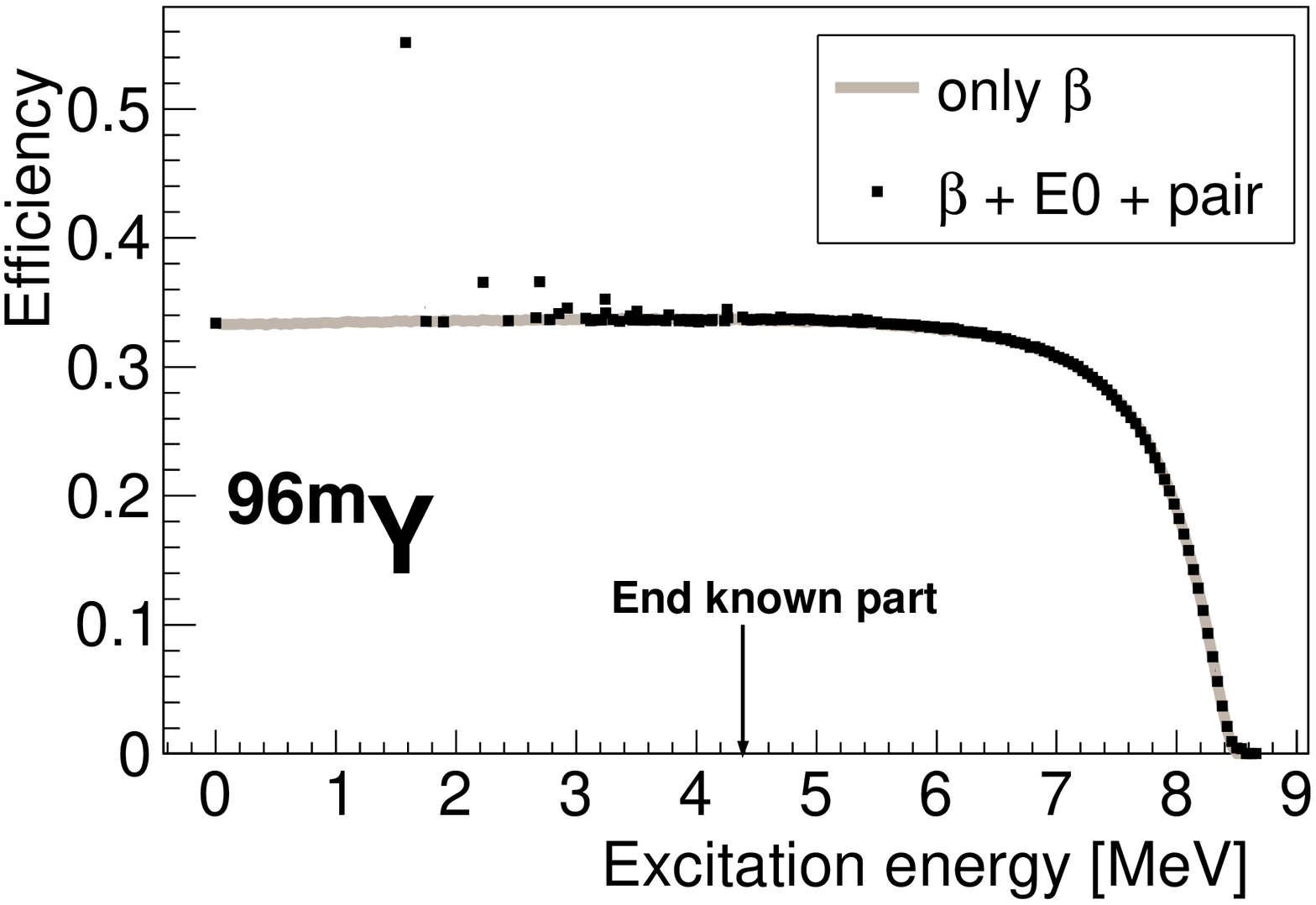}

\caption{Efficiency of the $\beta$ detector as a function of the excitation energy in the daughter nucleus ($^{96}$Zr). The efficiency considering only $\beta$ particles (grey line) is compared with that obtained taking conversion electrons and pair production into account as well (black squares). The upper and lower panels show the comparisons for $^{96\text{gs}}$Y and $^{96\text{m}}$Y, respectively. See text for more details. }
\label{beta_eff}
\end{center}
\end{figure}

Apart from drastically modifying the MC response for this level, taking into account the effect of conversion electrons and pair production also modifies the efficiency of the plastic $\beta$ detector associated with the decay to each level of the daughter nucleus that de-excites through the 1581.6~keV level, which is employed to normalize the MC responses in $\beta$-gated TAGS spectra. One expects an efficiency increase due to conversion electrons (and electrons from the pair) with respect to a situation where only $\beta$ electrons are emitted. Note that the sensitivity of the $\beta$ detector to $\gamma$ rays is very small and this effect is neglected. This change in the efficiency affects not only the response to the de-excitation of the 1581.6~keV level, but also that of all those levels de-exciting through it. In order to correct the $\beta$ efficiency of the MC simulations, we have introduced a novel approach that will be further exploited in Sec.~\ref{Mm_ana}. By means of a modified DECAYGEN event generator~\cite{TAS_decaygen} we have constructed an event file for each of the levels of $^{96}$Zr. The DECAYGEN program uses the branching ratio matrix as input and it has been modified to include the conversion electrons and the pair production. For each level in $^{96}$Zr we have performed a MC simulation with the corresponding event file as input and we have determined the efficiency of the plastic $\beta$ detector. In Fig.~\ref{beta_eff} we present a comparison between the $\beta$ efficiency of the plastic detector when only $\beta$ particles are considered and the one that also takes into account conversion electrons and pair production. 

As expected, the most important effect in the $\beta$-detector efficiency is observed for the 1581.6~keV level (with an increase from 34\% to 58\%), but a noticeable effect is also seen for those levels in the known part of the level scheme of $^{96}$Zr that are connected with it. In addition, since we consider 0$^-$ and 1$^-$ levels in the continuum region of $^{96}$Zr for the decay of the 0$^{-}$ ground state of $^{96}$Y (i.e., allowed $\beta$ transitions, including unlikely 0$^-\rightarrow$0$^-$ Fermi transitions), the efficiency for those levels is also affected due to their likely connection with the 0$^+$ 1581.6~keV level (see enhanced $\beta$ efficiency at high excitation energies in Fig.~\ref{beta_eff} top). On the contrary, for the decay of the 8$^{+}$ isomer allowed $\beta$ transitions to 7$^+$, 8$^+$ and 9$^+$ levels in the continuum region have been considered, and they de-excite through $\gamma$ cascades that do not pass through the 0$^+$ 1581.6~keV level (see $\beta$ efficiency at high excitation energies in Fig.~\ref{beta_eff} bottom).

Note that this is the first time that the decay through pair production and conversion electrons is taken into account in published TAGS analyses, allowing an independent determination of the $\beta$ feeding of this level from previous measurements. Indeed in a recent independent measurement with the TAGS technique of the decay of $^{96\text{gs}}$Y~\cite{MTAS_neutrino_PRL,MTAS_Acta_96Y} the authors do not consider the pair production in their response (although the 1022~keV peak is clearly seen in figure 1 from Ref.~\cite{MTAS_Acta_96Y}). They also do not consider conversion electrons, and their MC response of the 1581.6~keV level is analogous to the response for the ground state. In other words they only consider $\beta$ electrons, as shown by the grey line in Fig.~\ref{1582_response}. The authors mention an "inefficiency of MTAS to clearly detect the conversion electrons"~\cite{MTAS_Acta_96Y} and they had to fix the feeding to the 1581.6~keV level to the value previously known (1.26$\%$~\cite{NDS_96}), thus omitting the effect of these conversion electrons on the efficiency of their $\beta$ detector. 

As a final comment, in our analyses we have neglected the pair production branch for other minor $E0$ transitions in $^{96}$Zr: the 1113.53 and 2695.17~keV transitions from the 0$^+$ level at 2695.18~keV and the 1343.89 and 2925.50~keV transitions from the 0$^+$ level at 2925.5~KeV. In contrast with the level at 1581.6~keV, these levels de-excite predominantly through $\gamma$ branches and not by conversion electrons. In addition, the probability of pair emission for those $E0$ branches cannot be calculated with BrIcc (v2.3) because either they are too close to the 2$m_{e}$ threshold energy (the 1113.53 and 1343.89~keV transitions) or they are outside the energy range of BrIcc (as it is the case of the 2695.17 and 2925.50~keV transitions). Note that together, those two levels are directly fed with 0.2\% probability in the decay of $^{96\text{gs}}$Y (and 0.02\% indirectly) according to ENSDF~\cite{NDS_96}, with no feeding in the case of the decay of $^{96\text{m}}$Y. 

\subsection{$^{96\text{gs}}$Y}\label{gs}

\begin{figure}[h]
\begin{center}
\includegraphics[width=0.5 \textwidth]{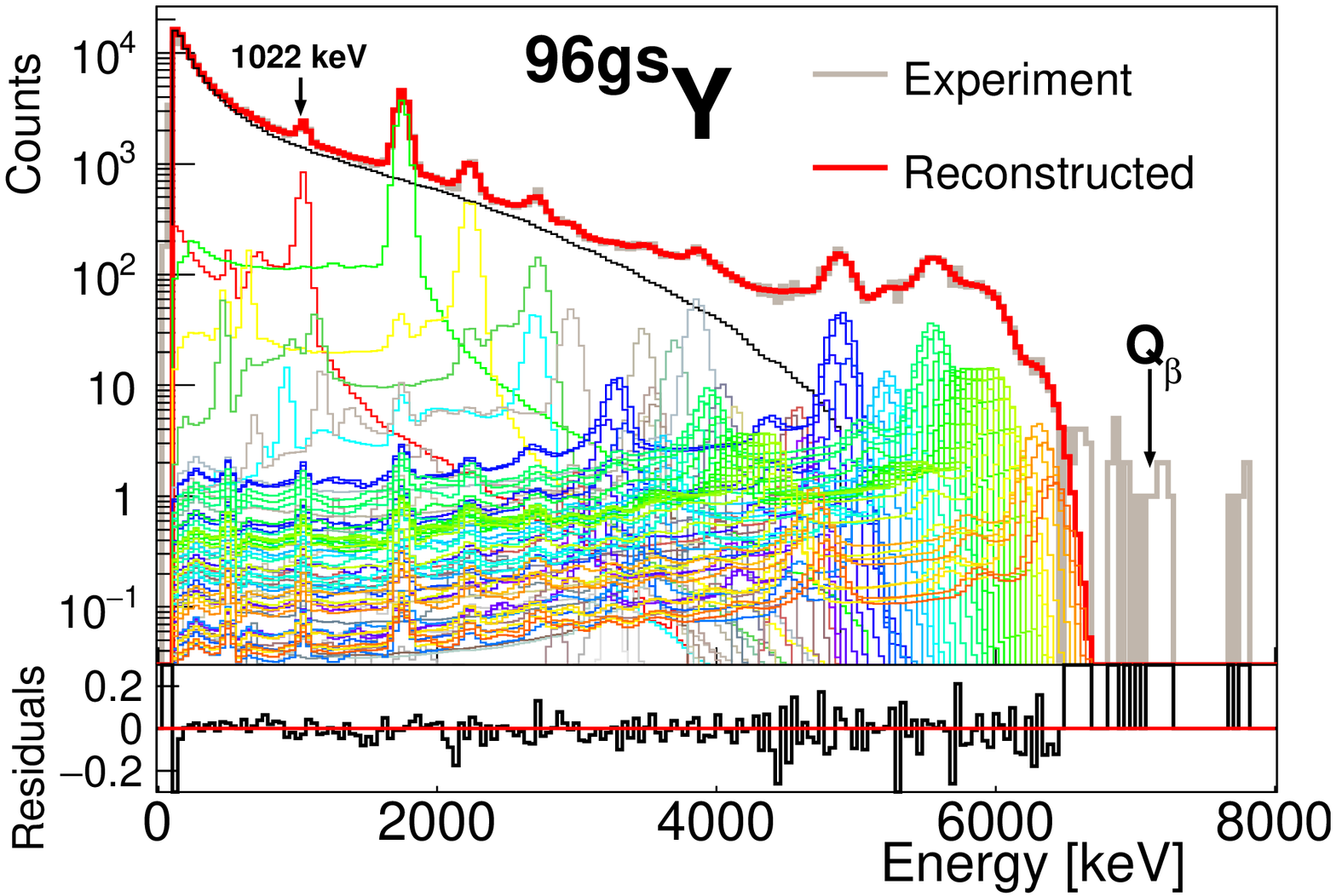} \\

\includegraphics[width=0.5 \textwidth]{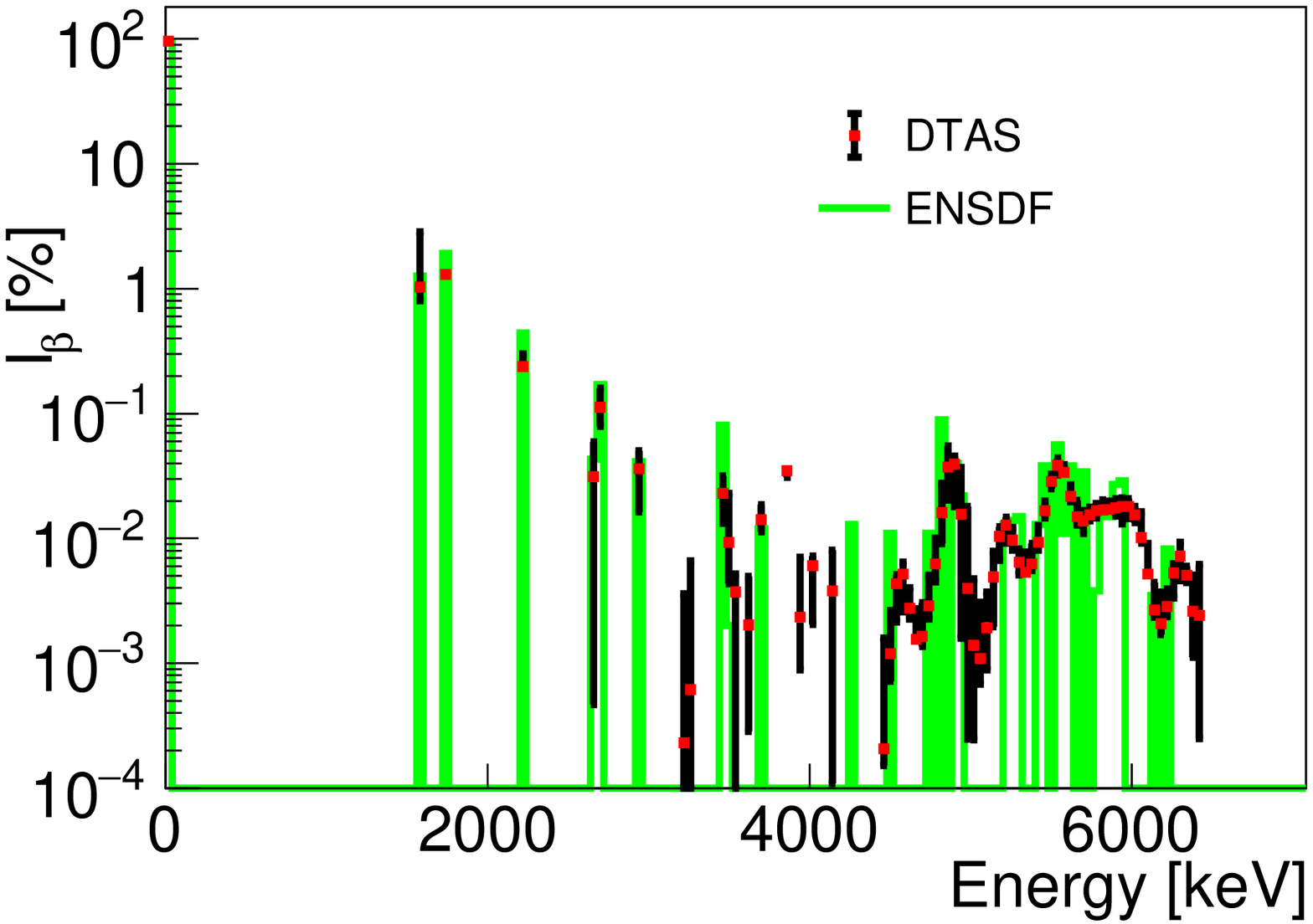}

\caption{Top panel: experimental $\beta$-gated spectrum summing-pileup subtracted for $^{96\text{gs}}$Y (grey) and reconstructed spectrum (red). The MC responses of each level fed in the daughter nucleus are shown with thinner lines and the 1022~keV peak due to the positron annihilation is highlighted. The relative deviations between experimental and reconstructed spectra are shown. Bottom panel: $\beta$ intensities for the present TAGS results (red dots with error bars) and high-resolution $\gamma$-spectroscopy data from ENSDF (green line).}
\label{96gsY_fit}
\end{center}
\end{figure}

For the analysis of the decay of $^{96\text{gs}}$Y we considered allowed transitions plus first forbidden transitions to levels in the known part of the level scheme of $^{96}$Zr, and only allowed transitions in the continuum part, as mentioned earlier. The quality of the TAGS analysis can be seen in the top panel of Fig.~\ref{96gsY_fit} by comparing the experimental spectrum with the spectrum reconstructed by the convolution of the $\beta$ intensities obtained in the analysis with the corresponding response function of the spectrometer. 

In the bottom panel of Fig.~\ref{96gsY_fit} the $\beta$-intensity distribution obtained in this work (supplied in the Supplemental Material~\cite{Suplement}) is compared with the high-resolution spectroscopy values from ENSDF~\cite{NDS_96} based on Ref.~\cite{Mach_96Y}. The decay is dominated by the ground state to ground state feeding intensity. We determine a value of $96.6_{-2.1}^{+0.3}\%$, compatible within uncertainties with the 95.5(5)$\%$ value quoted in ENSDF based on Ref.~\cite{Mach_96Y} (equal to the 95.5(20)$\%$ value determined in the recent TAGS study previously mentioned~\cite{MTAS_neutrino_PRL}). The present value is found to be affected by the change in the efficiency due to conversion electrons discussed in the previous section, and a 95.7$\%$ value is obtained without applying this correction. In addition, if we only take the response to $\beta$ particles for the level at 1581.6~keV excitation energy (see grey line in Fig.~\ref{1582_response}), the $\beta$ intensity is shared between the ground state (74.7$\%$) and the 0$^+$ level at 1581.6~keV (22.9$\%$), since the MC responses for both levels are very similar. In order to get reasonable results, we need to fix the $\beta$ intensity to the 1581.6~keV level to the 1.26$\%$ value from ENSDF~\cite{NDS_96}, as done in Ref.~\cite{MTAS_neutrino_PRL}, thus obtaining a value of 96.4$\%$. We have not taken into account such a solution with only $\beta$ particles in the MC response of the level at 1581.6~keV, because it does not reproduce the clear 1022~keV peak seen in our experimental spectrum (as highlighted in Figs.~\ref{96Y_spectra} and \ref{96gsY_fit}). Our analysis with the correct response function naturally gives a $1.03_{-0.23}^{+1.83}$\% $\beta$ intensity to the 0$^+$ level at 1581.6~keV, in agreement within the uncertainties with the 1.26(10)\% ENSDF value. On the other hand, the effect of the first bin included in the analysis (with each bin equivalent to 40~keV) was also found to affect the value of the ground state feeding probability by up to $\pm$0.4$\%$. Similarly, we have observed that the ground state feeding intensity value is very sensitive to variations of the $P_{e^-e^+}$ value: a $\pm$50$\%$ variation has been observed to change the ground state feeding intensity to 97.0$\%$ and 95.5$\%$, respectively. However, such a large error in $P_{e^-e^+}$ is not justified and we have considered a $\pm$25$\%$ variation in the estimation of the uncertainties, as discussed in Section~\ref{1581}. Finally, we have also applied the $4\pi\gamma-\beta$ counting method introduced by Greenwood \textit{et al.}~\cite{Greenwood_GS} and recently revisited in Ref.~\cite{PRC_GS}. It is based on the number of counts registered in the DTAS $\beta$-gated spectrum together with the number of counts registered in the $\beta$ detector. We obtain a value of $93(3)\%$, compatible within uncertainties with the TAGS result. The large uncertainty is due to the influence of the conversion electrons both in the efficiencies and in the number of counts from DTAS and the plastic $\beta$ detector. 

Different sources of systematic error have been considered that may contribute to the uncertainties of the $\beta$ intensities in this work. The statistical errors are negligible in comparison. Solutions without correcting the $\beta$-detector efficiency have been included in the error budget, as well as changing the $P_{e^-e^+}$ value by up to $\pm$25$\%$. The different spin-parity values discussed above for the branching ratio matrix were considered for the estimation of the uncertainties, and resulted in a negligible impact on the $\beta$-intensity distribution. In addition, two possible sets of correction factors for the HFB+c level density distribution were employed (the one available at RIPL-3 and another set of correction factors that slightly improves the reproduction of the accumulated number of known levels in $^{96}$Zr at low excitation energies). The difference between the two sets of corrections was also negligible. Similarly, the two possible sets of PSF parameters for $E1$ transitions presented in Table~\ref{parameters_PSF} gave equivalent results in the analysis.

We also modified the normalization of the summing-pileup by $\pm$50$\%$ until the reproduction of the experimental spectrum with the result of the analysis was not acceptable. Due to the large ground state to ground state branch, random coincidences between $\beta$ particles detected in the $\beta$ detector and environmental $\gamma$ background in DTAS may be enhanced~\cite{100Tc}. We have considered the influence in our analysis of subtracting an environmental background component without noticing any significant difference. The possible influence of the deconvolution algorithm has also been evaluated by employing the Maximum-Entropy algorithm in addition to the Expectation-Maximization one. Finally, errors in the energy and resolution calibrations have been considered in the uncertainties seen in the bottom panel of Fig.~\ref{96gsY_fit}, as well as the effect of the threshold of the $\beta$ detector (which affects the energy dependence of the efficiency of this detector).

\begin{figure*}[t]
\begin{center}
\includegraphics[width=1 \textwidth]{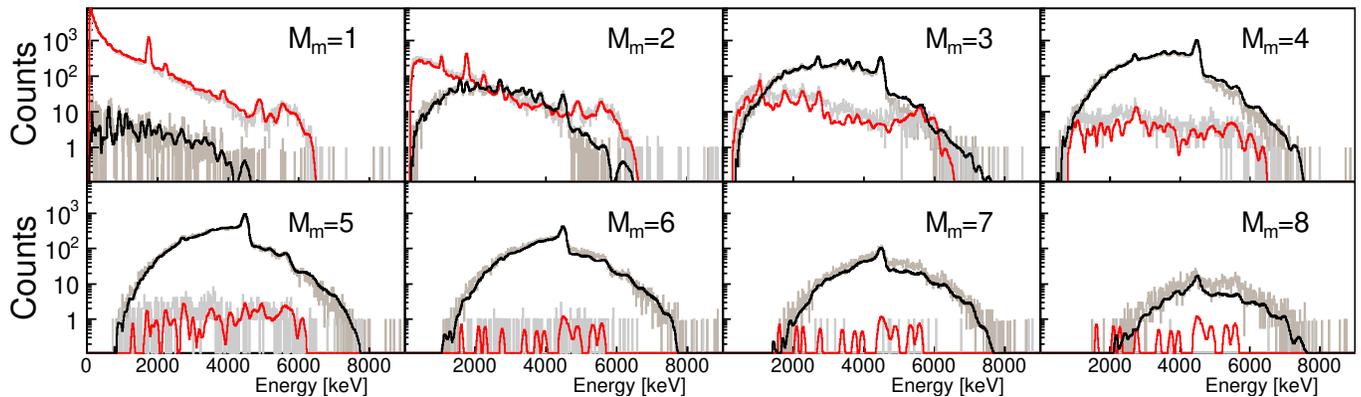}
\caption{Module-multiplicity gated TAGS spectra for $M_m$=1-8. The experimental spectra free of contaminants and the MC spectra obtained with the results of the TAGS analysis are compared for the decay of $^{96\text{gs}}$Y (experiment: light grey, MC: red) and the decay of $^{96\text{m}}$Y (experiment: grey, MC: black).}
\label{Mm_all}
\end{center}
\end{figure*}

As a cross-check of our branching ratio matrix and our response function, we have investigated the reproduction of other experimental observables with the results of the reference analysis, in line with our recent works~\cite{PRC_Nb,PRC_BDN}. For this the previously mentioned DECAYGEN event generator has been employed, with the branching ratio matrix and the $\beta$-intensity probabilities of the TAGS analysis as input. It allowed us to study the reproduction of the spectra of the individual modules of DTAS, as well as the total absorption spectra with module-multiplicity ($M_m$) conditions (where $M_m$ represents the number of modules of DTAS that register a signal above the threshold for a given event). In both cases we found a nice reproduction of the experimental spectra, as shown in Fig.~\ref{Mm_all} for the $M_m$-gated spectra.

We have also checked the reproduction of the absolute intensities of the strongest $\gamma$ transitions of the decay, corresponding to the 2$^+$ level at 1750~keV excitation energy in $^{96}$Zr. Our reference analysis gives a value of 0.016 and a solution obtained with a branching ratio matrix modified to reproduce the 0.024 value of this $\gamma$ intensity quoted in ENSDF~\cite{NDS_96} has also been tried. However, the latter value worsened the reproduction of the experimental total absorption spectrum and it was not included in the error budget. 

\subsection{$^{96\text{m}}$Y}\label{m}

\begin{figure}[h]
\begin{center}
\includegraphics[width=0.5 \textwidth]{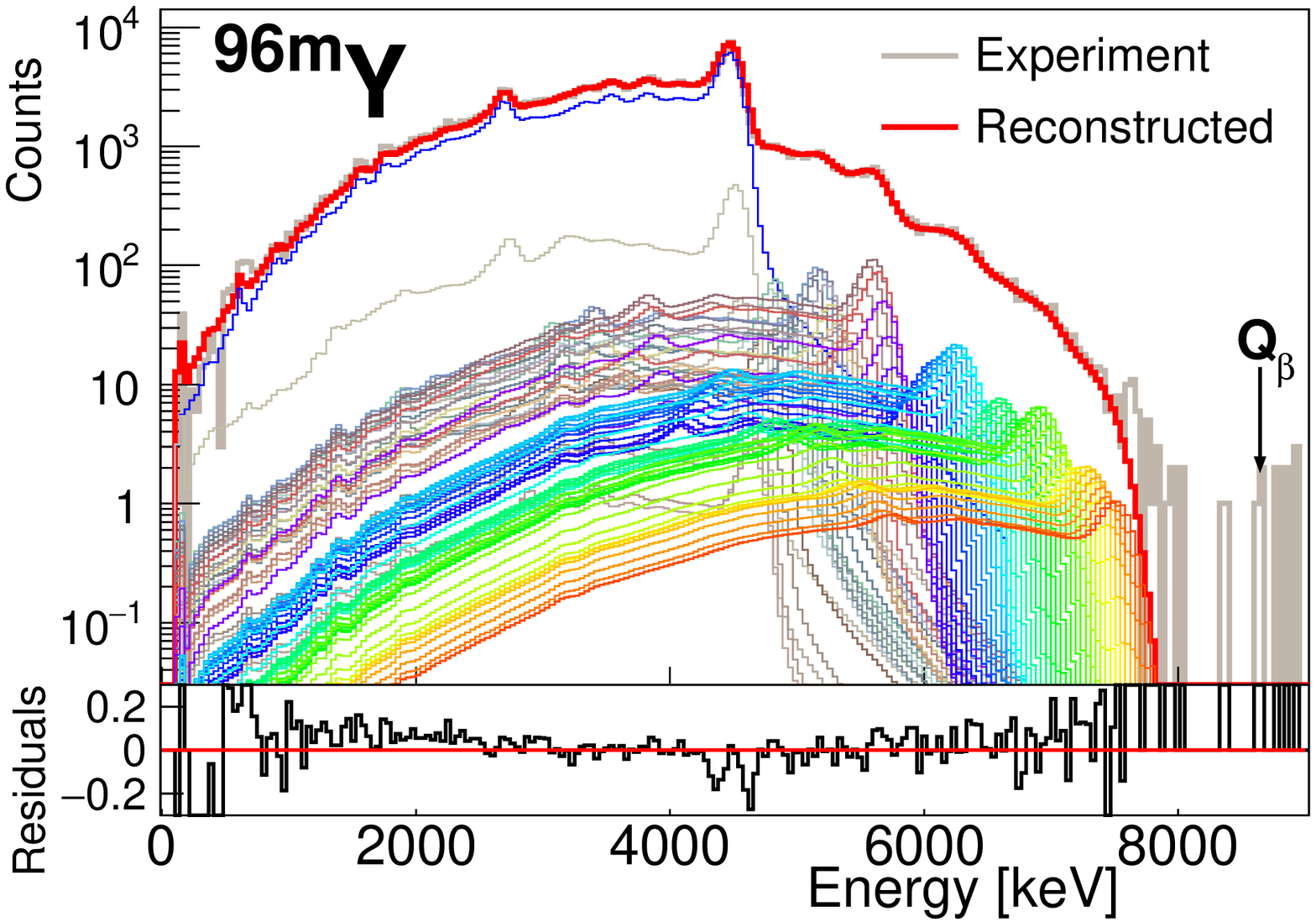} \\

\includegraphics[width=0.5 \textwidth]{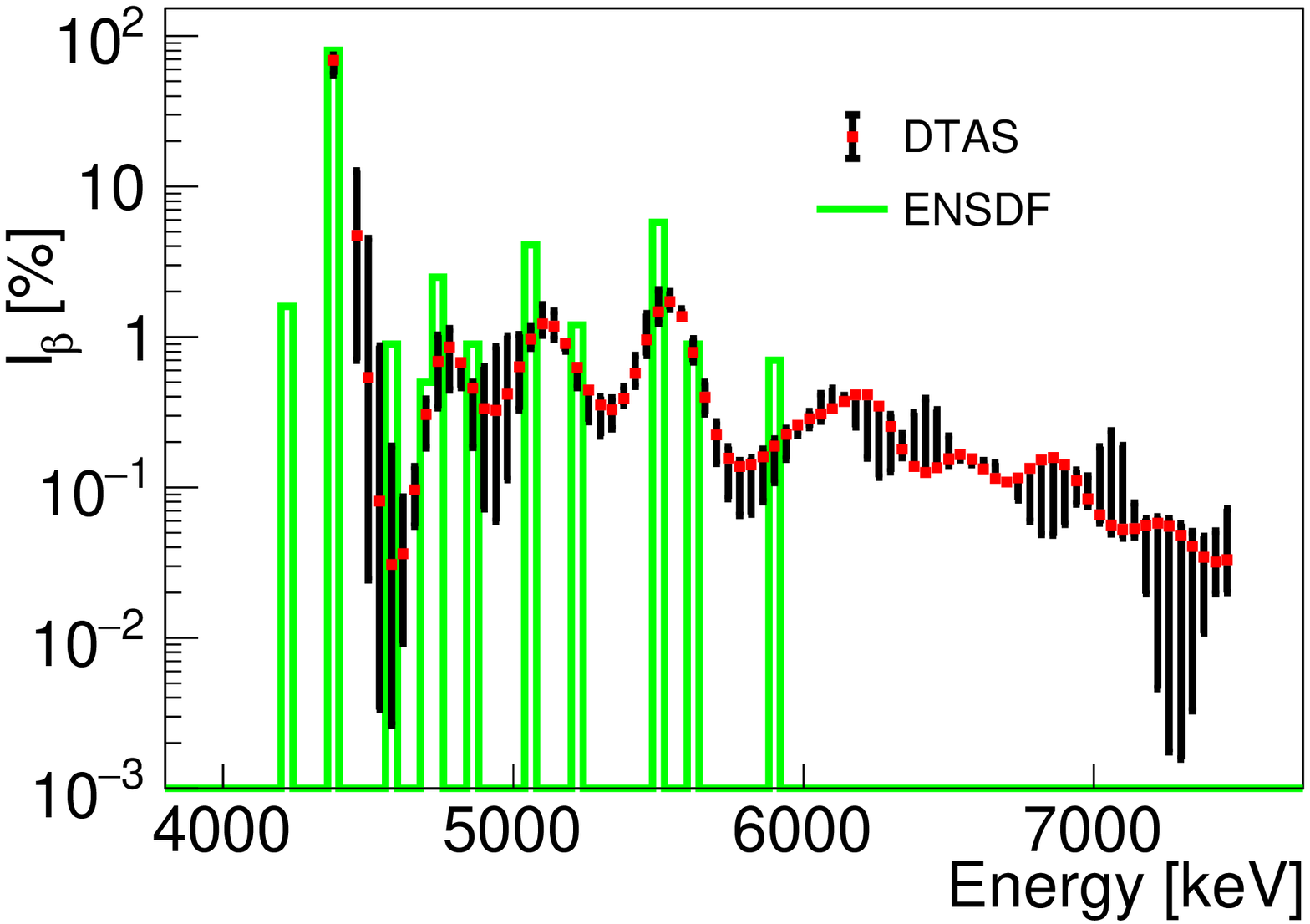}

\caption{Top panel: experimental $\beta$-gated spectrum summing-pileup subtracted for $^{96\text{m}}$Y (grey) and reconstructed spectrum (red). The MC responses of each level fed in the daughter nucleus are shown with thinner lines. The dominant level fed is shown as a blue line. The relative deviations between experimental and reconstructed spectra are shown. Bottom panel: $\beta$ intensity for the present TAGS results (red dots with error bars) and high-resolution $\gamma$-spectroscopy data from ENSDF (green line).}
\label{96mY_fit}
\end{center}
\end{figure}

For the TAGS analysis of the decay of the 8$^{+}$ isomeric state, direct feeding to 7$^+$, 8$^+$ and 9$^+$ levels in $^{96}$Zr (allowed transitions) was considered both in the known part of the level scheme and in the continuum region. We thus avoid direct feeding to the 7$^-$ level at 4234.7~keV excitation energy, populated with 1.6\% $\beta$ intensity according to previous high-resolution studies~\cite{Julich_96mY}, since it did not to affect the quality of the reproduction of the experimental spectrum. In the top panel of Fig.~\ref{96mY_fit} we show the reproduction of the experimental spectrum with the results of the TAGS analysis. 

In this case it was possible to modify the branching ratio matrix to achieve an improved reproduction of the known $\gamma$ intensities $I_{\gamma}$ of the low-energy levels without deteriorating the reproduction of the experimental spectrum. With regard to the normalization of the $I_{\gamma}$ to absolute intensities per 100 decays, the 0.088 factor of the ENSDF evaluation~\cite{NDS_96} was used. Note that with such a factor, however, the $I_{\beta}$ obtained from $I_{\gamma}$ balance are not those quoted in ENSDF (based on Ref.~\cite{Julich_96mY}), as explained in the ENSDF evaluation~\cite{NDS_96}. This is due to the fact that in Ref.~\cite{Julich_96mY} $I_{\gamma+ce}$ of the 1581.6~keV $E0$ transition was neglected, even though the 0$^+$ level at 1581.6~keV was found to be indirectly populated from a 2$^+$ level at 2226.2~keV. In Table~\ref{Igamma} the $\gamma$ intensities from ENSDF~\cite{NDS_96} and those obtained in the TAGS analysis with and without the modified branching ratio matrix are presented. The $\beta$-intensity distribution determined with the modified branching ratio matrix was considered inside the error budget as an alternative to the reference one shown in the lower panel of Fig.~\ref{96mY_fit}.

\begin{table}[h]
\begin{ruledtabular}
\centering
\begin{tabular}{cccc}
Energy [keV] & $I_{\gamma}$ ENSDF & $I_{\gamma}$ DTAS & $I_{\gamma}$ DTAS$^*$ \\ \hline
  1750.5   & 0.88       &  0.88   &    0.87   \\
  1897.2   & 0.39    &  0.47   &    0.41   \\
  2225.8   & 0.11    &  0.08   &    0.09   \\
  2857.4   & 0.60    &  0.49   &    0.60   \\
  3119.9   & 0.27    &  0.38   &    0.32   \\
  3483.4   & 0.26    &  0.22   &    0.25   \\
  3772.2   & 0.63     &  0.51   &    0.63   \\
  4389.5   & 0.76     &  0.69   &    0.75   \\
\end{tabular}
\caption{\label{Igamma} Absolute $\gamma$ intensities per 100 decays de-exciting the main levels in the known part of the level scheme populated in the decay of $^{96\text{m}}$Y. The second column corresponds to the intensities obtained from high-resolution $\gamma$-ray spectroscopy studies~\cite{NDS_96}. The third column gives the intensities obtained with DTAS for the reference analysis, whereas the intensities obtained with a modified branching ratio matrix are presented in the fourth column (DTAS$^{*}$).}
\end{ruledtabular}
\end{table}

We have investigated the same sources of systematic uncertainty mentioned above in the analysis of the $^{96\text{gs}}$Y. In this case the normalization factor of the summing-pileup component could be changed by up to $\pm$80$\%$, whilst still obtaining a good reproduction of the experimental spectrum. Likewise, the parent activity normalization factor and the normalization factor of the $^{96\text{gs}}$Y contamination have been changed by $\pm$10$\%$. All these sources of uncertainty define the error bars of the $I_{\beta}$ distribution presented in Fig.~\ref{96mY_fit} bottom (available in the Supplemental Material~\cite{Suplement}), where the $\beta$ feedings from the ENSDF evaluation~\cite{NDS_96} are shown for comparison. A sizable Pandemonium effect is observed in the ENSDF data, based on high-resolution results~\cite{Julich_96mY}. In particular, we obtain 6\% of the $\beta$ intensity above 5900.1~keV excitation energy, the last level in $^{96}$Zr previously known to be populated in the $\beta$ decay of the 8$^{+}$ isomer~\cite{Julich_96mY}. Note that recent beyond-mean field calculations~\cite{Petro96Y_2020} predict significant $\beta$ strength associated with allowed transitions to 7$^{+}$, 8$^{+}$ and 9$^{+}$ states in this last part of the $\beta$-energy window, where no experimental data were available until now.  

As in the decay of the ground state, we have investigated the reproduction of the $M_m$-gated spectra and the individual spectra with the results of the analysis of the decay of $^{96\text{m}}$Y. Good agreement was found within statistical uncertainties, as shown in Fig.~\ref{Mm_all} for the TAGS spectra gated in $M_m$ from 1 to 8.

\subsection{Analysis of the $M_m$-gated spectra}\label{Mm_ana}

Due to the different spin-parity values of $^{96\text{gs}}$Y (0$^-$) and $^{96\text{m}}$Y (8$^+$), the decay patterns are found to be quite dissimilar, as shown in Fig.~\ref{96Y_spectra}. It also implies a difference in the $\gamma$ multiplicity ($M_{\gamma}$) of the cascades de-exciting the levels populated in $\beta$ decay, which translates, in turn, into differences in the experimental module-multiplicity $M_m$ spectra. As can be seen in the study of the reproduction of the $M_m$-gated spectra of Fig.~\ref{Mm_all}, the decay of $^{96\text{gs}}$Y favors low $M_m$ spectra ($M_m$=1-2 dominate), while the decay of $^{96\text{m}}$Y preferentially favors high $M_m$ spectra ($M_m$=3-6 dominate in this case). This suggests the possibility of studying the $\beta$ decay of the 8$^{+}$ isomer by looking at high $M_m$ spectra in a combined measurement of both $\beta$-decaying states. It would be a useful strategy for other cases with decaying isomers lying very close in energy to the ground states and such a strong difference in spin-parity values, for which an assisted-trap separation cannot be achieved. It would also allow us to implant a mixture of the decaying states instead of applying extra purification techniques that normally significantly lower the implantation rate. For the present case we can explore this innovative possibility thanks to some runs where we implanted $^{96\text{m}}$Y with a contamination of $^{96\text{gs}}$Y.

\begin{figure}[h]
\begin{center}
\includegraphics[width=0.5 \textwidth]{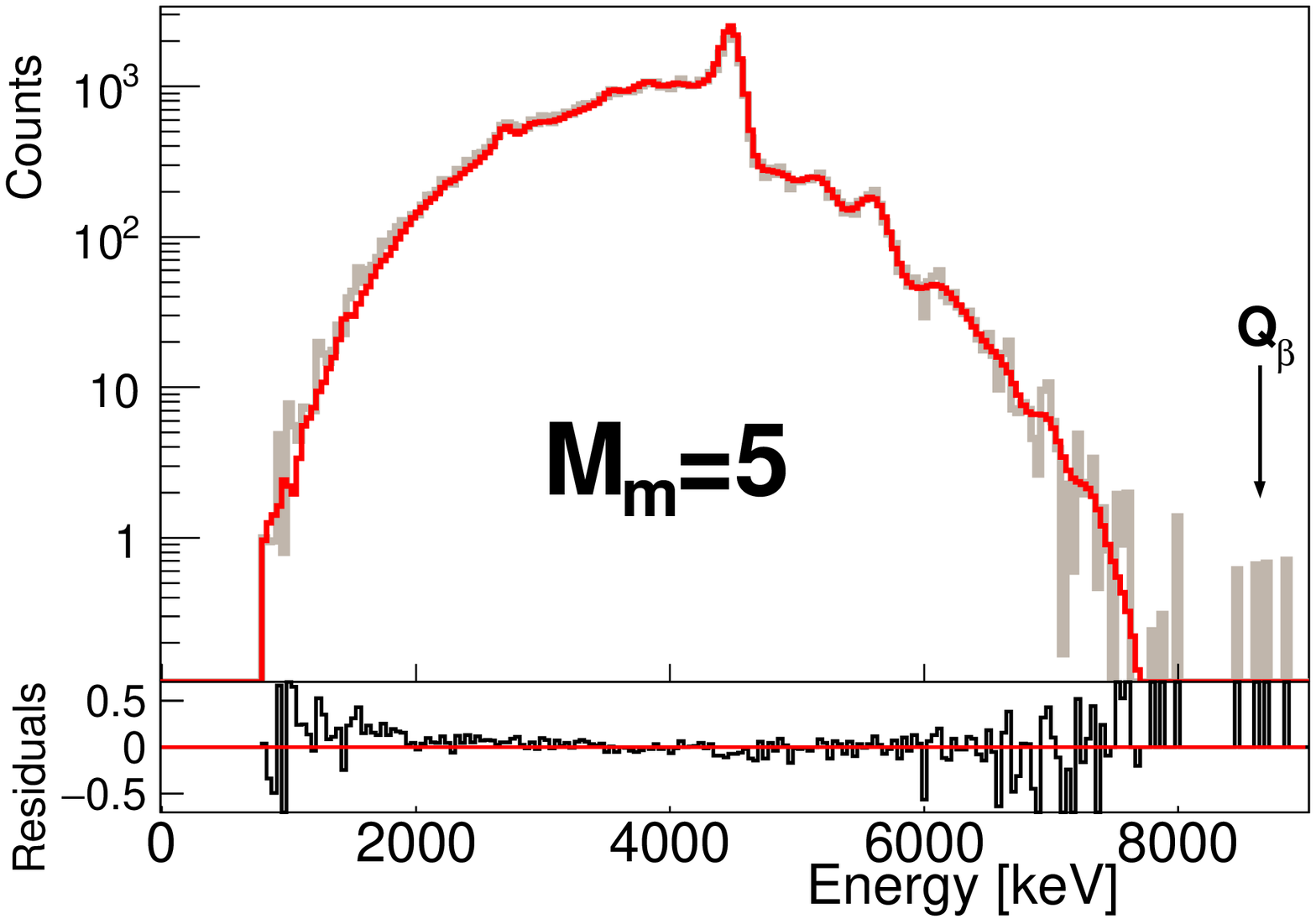} \\

\includegraphics[width=0.5 \textwidth]{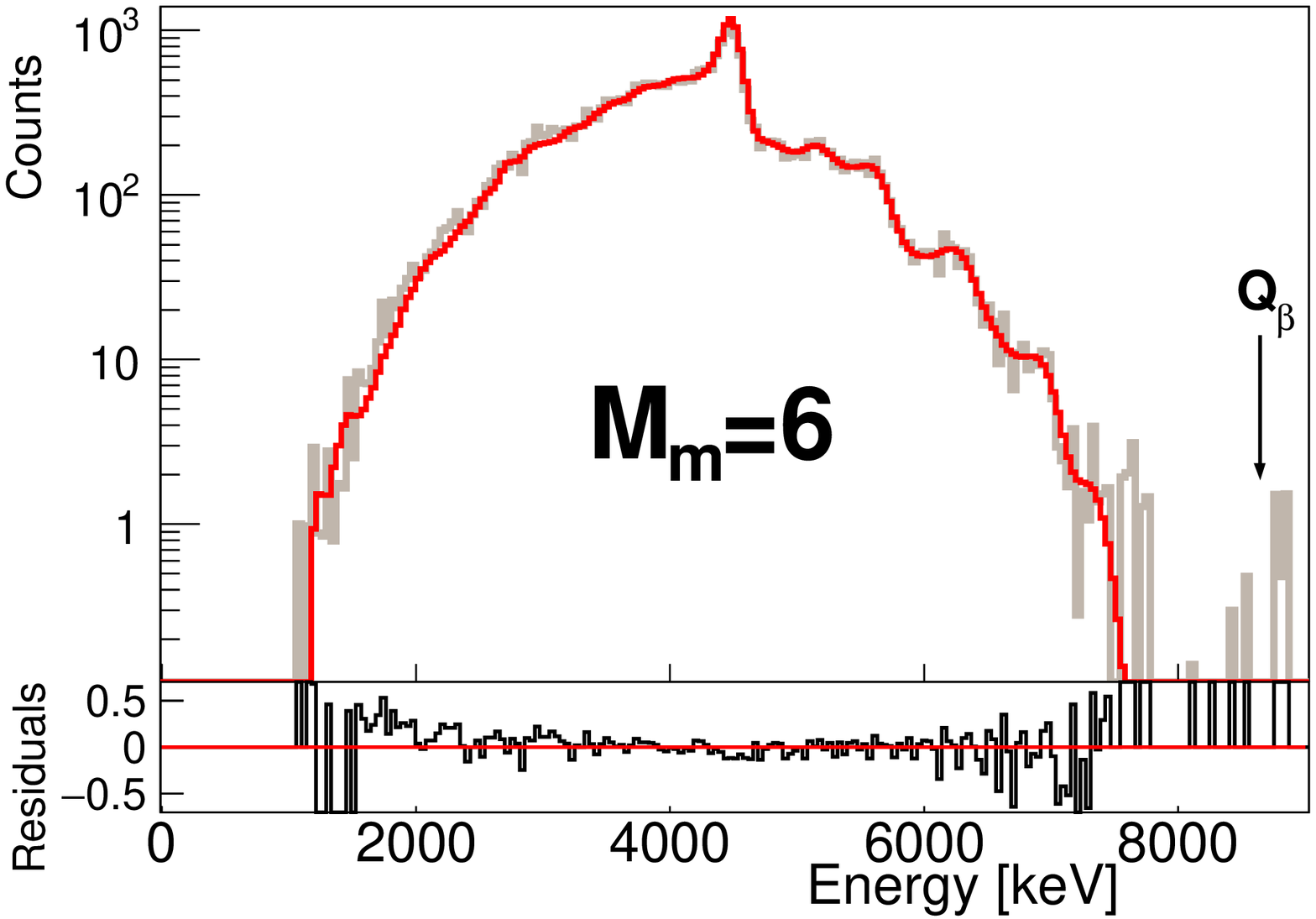}

\caption{Experimental $\beta$-gated TAGS spectra (grey line) with conditions in $M_m$=5 (top) and $M_m$=6 (bottom) for a measurement of $^{96\text{m}}$Y with a contamination of $^{96\text{gs}}$Y. The experimental spectra are summing-pileup subtracted. The reconstructed spectra with the results of the corresponding TAGS analyses are shown by the red line. The relative deviations between experimental and reconstructed spectra are shown in both cases.}
\label{96mY_fit_Mm}
\end{center}
\end{figure}

\begin{figure}[h]
\begin{center}
\includegraphics[width=0.5 \textwidth]{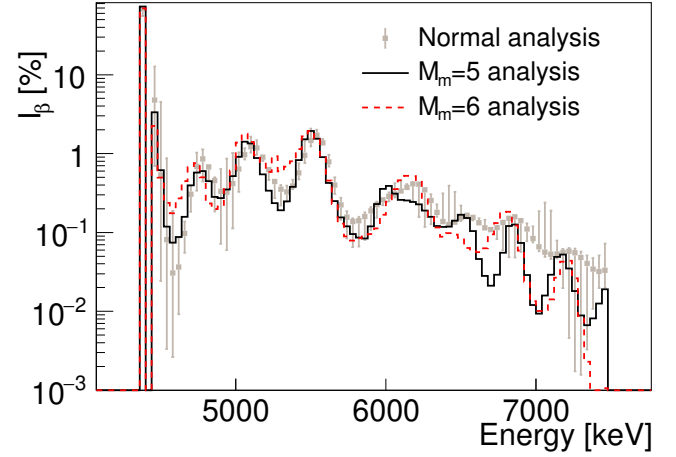}

\caption{$\beta$-intensity distribution determined in the normal TAGS analysis (grey dots with error bars) compared with those obtained in the TAGS analyses of the $M_m$=5 (solid black) and $M_m$=6 (dotted red) TAGS spectra.}
\label{96mY_I_Mm}
\end{center}
\end{figure}

Regarding the TAGS analysis of the $M_m$-gated spectra, we have applied the same strategy presented in Section~\ref{1581}, i.e. to use a modified DECAYGEN event generator~\cite{TAS_decaygen} to construct an event file for each of the levels of $^{96}$Zr. The MC simulations with such event files are then employed to construct the response function of the spectrometer for each $M_m$, using the same branching ratio matrix employed for the normal analysis. Recently a similar approach was used to construct the response function for the decay of $^{186}$Hg~\cite{TAS_186Hg}, in order to take properly into account a summing effect with X-rays. The $\beta$-intensity distribution is determined for each $M_m$ spectrum with the EM algorithm~\cite{TAS_algorithms}. We have applied this new method to the high module-multiplicity spectra of the runs where the two decaying components, $^{96\text{gs,m}}$Y, were implanted together. As shown in Fig.~\ref{Mm_all}, the decay of the ground state hardly produces events of $M_m>$4, and we can thus treat these spectra as coming only from the decay of the 8$^{+}$ isomer. We have performed the TAGS analyses of the $M_m$=5, 6 spectra, and the quality of the reproduction of the experimental data with the results of the analyses is shown in Fig.~\ref{96mY_fit_Mm} after considering the corresponding summing-pileup contribution. In Fig.~\ref{96mY_I_Mm} the $\beta$-intensity distributions determined in these TAGS analyses are compared with the reference $\beta$ feedings obtained in Sec.~\ref{m} for the total spectrum. A reasonable agreement is found, proving the validity of this method to determine $\beta$-intensity distributions with the TAGS technique from the $M_m$ spectra. In a second step, one should employ the $\beta$-intensity distribution determined from the high $M_m$ spectra (for example from the $M_m$=5 spectrum, the one with more statistics) in a MC simulation using the DECAYGEN event generator. The results of these simulations should be used to subtract the decay of the 8$^{+}$ isomer from the low $M_m$ spectra of the measurement with the two decaying components in order to isolate the $^{96\text{gs}}$Y one. Unfortunately, in this particular case we could not obtain useful $^{96\text{gs}}$Y $M_m$-gated spectra in this way, due to limited statistics in the low $M_m$ spectra of the measurement with the two decaying components. This is due to the fact that we were accidentally implanting some $^{96\text{gs}}$Y in a measurement of $^{96\text{m}}$Y and not intentionally a real mixture of both $\beta$-decaying states, which in this case are almost equally produced, as shown in the mass scan of Fig.~\ref{96Y_mass} and in accordance with previous proton-induced fission yield measurements at IGISOL~\cite{FY_96Y_IGISOL}.  

\subsection{Reactor summation calculations}\label{Reactor}

\begin{figure}[h]
\begin{center}
\includegraphics[width=0.5 \textwidth]{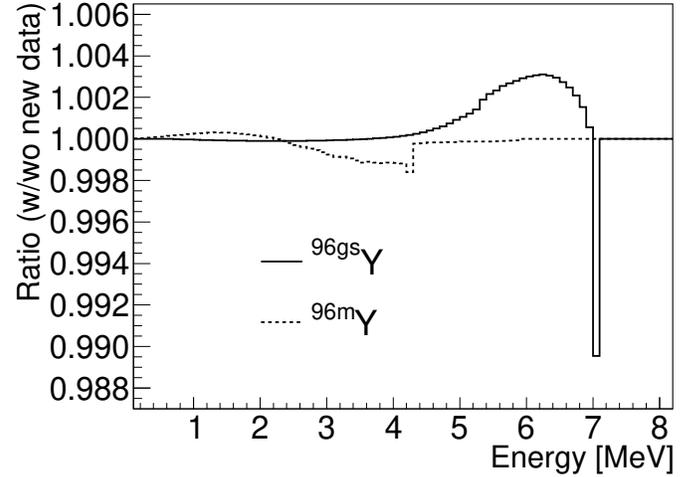}
\caption{Ratio of reactor antineutrino spectra for $^{235}$U, with and without the present new data, as a function of energy when the results obtained in the present work replace Rudstam's~\cite{Rudstam} data for $^{96\text{gs}}$Y and JEFF-3.3 data for $^{96\text{m}}$Y. The effect of $^{96\text{gs}}$Y (solid line) and $^{96\text{m}}$Y (dotted line) are presented separately. The spike observed for $^{96\text{gs}}$Y is due to a difference in the $Q_{\beta}$ value between the original calculation and the current value.}
\label{235U}
\end{center}
\end{figure}

As mentioned in the introduction, the decay of $^{96\text{gs}}$Y is one of the most important contributors to the reactor antineutrino spectrum in the high-energy region. The summation method developed by the group of Nantes~\cite{neutrinos_PRL,PRL_Magali} has been employed to study the impact of the present TAGS results for each of the four main fissile isotopes in a PWR: $^{235}$U, $^{239}$Pu, $^{241}$Pu, and $^{238}$U. Until now, in the Nantes summation method the data from the Joint Evaluated Fission and Fusion File (JEFF-3.3) database~\cite{JEFF33} were used as input for the decay of $^{96\text{m}}$Y, while the data from Rudstam et al.~\cite{Rudstam} were used  for $^{96\text{gs}}$Y. The latter are based on the $\beta$-spectra measurements performed by Tengblad \textit{et al.} at OSIRIS-ISOLDE with another Pandemonium-free technique that employed a $\beta$ spectrometer~\cite{Olof_beta}. The impact of replacing those data by our TAGS results was found to be small. As shown in Fig.~\ref{235U} for $^{235}$U, a difference below 0.5$\%$ is obtained in the ratio between the new summation calculation and the one with previous data. Similar figures are obtained for the other three fissile isotopes, with the largest impact in the region of the antineutrino spectral shape distortion between 5 and 7~MeV. The reason for this modest impact is twofold: on the one hand because of the similar $\beta$ intensities obtained with respect to previous measurements for $^{96\text{gs}}$Y, in particular for the g.s.\ feeding probability, the dominant branch of the decay, and on the other hand because of the low cumulative fission yield of $^{96\text{m}}$Y, which amounts to 0.011(2) for $^{235}$U in comparison with 0.047(2) for the ground state~\cite{JEFF33}. The confirmation of the role of the decay of $^{96\text{gs}}$Y with the present results is specially important given that it is one of the decays contributing more in the region of the spectral shape distortion, adding more than 12$\%$ of the antineutrino spectrum of a PWR in the 5-7~MeV energy range~\cite{Zak_PRL}.

We have evaluated the average $\gamma$ and $\beta$ energies obtained with the present TAGS results. Due to the strong $E0$ transition of the level at 1581.6~keV excitation energy in $^{96}$Zr, we have separately evaluated the mean energy of the conversion electrons (as well as the average X-ray energy and the average annihilation energy). The average energies are listed in Table~\ref{Mean_exp} and for comparison we present the corresponding average energies from two decay databases: Evaluated Nuclear Data File (ENDF/B-VII.1)~\cite{ENDF} and JEFF-3.3~\cite{JEFF33}. In all cases, the energy of the corresponding X-rays has been combined with the mean $\gamma$ energy, as well as the energy coming from the annihilation of the positron when pair production competes with electron conversion. The errors associated with the present results correspond to the evaluation of the average energies for all the solutions used for the estimation of the uncertainties of the $\beta$ intensities presented in the lower panels of Figs.~\ref{96gsY_fit} and \ref{96mY_fit}. Large asymmetric error bars are quoted for $^{96\text{gs}}$Y, specially for the average conversion electron energy, due to the influence of the three main factors affecting the ground state feeding value mentioned before: (1) using the modified efficiency discussed in Sec.~\ref{1581} decreases the $\gamma$ and conversion electron mean energies in comparison to employing the original one, (2) decreasing the BrIcc factor has the opposite effect, increasing both $\gamma$ and conversion electron average energies, and (3) reducing by one unit the first bin considered in the analysis also increases $\gamma$ and conversion electron average energies. 

The newly obtained average $\beta$ and $\gamma$ energies for $^{96\text{gs}}$Y are close to the JEFF-3.3~\cite{JEFF33} and ENDF/B-VII.1~\cite{ENDF} values quoted in Table~\ref{Mean_exp}. As we mentioned before, this is due to the fact that, for this case, the $\beta$ intensities obtained in the present work are similar to previous results obtained from high-resolution $\gamma$-spectroscopy measurements. Nevertheless the uncertainties are reduced, especially in the case of the average $\beta$ energy. With $^{96\text{gs}}$Y contributing importantly to the DH at short cooling times, we expect that these reduced uncertainties impact the future uncertainty calculations on DH.

The average values obtained in the case of $^{96\text{m}}$Y reflect the Pandemonium effect found in the $\beta$-feeding distribution. The average $\gamma$ energy is larger by about 200~keV to more than 300~keV with respect to the JEFF-3.3 and ENDF/B-VII.1 databases. The average $\beta$ energy is lower than the value quoted in JEFF-3.3 but larger than in ENDF/B-VII.1 due to the different excitation energy attributed to the 8$^+$ isomer in ENDF/B-VII.1.

The average $\gamma$ and $\beta$ energies were used as input for decay heat summation calculations developed by the group of Nantes~\cite{DH_Lydie}, performed with the SERPENT2~\cite{Serpent} reactor burnup MC code coupled to the JEFF-3.3~\cite{JEFF33} decay data library, which was used as a reference for the calculations. As in the case of antineutrino spectrum summation calculations, the impact of the new results was found to be very small, both for the electromagnetic component (that accounts for $\gamma$ rays, X-rays and anhiliation) and the light particle component ($\beta$ electrons and conversion electrons), with a ratio below 0.5$\%$ between the calculations that include the present TAGS results and the ones with the reference data from JEFF-3.3.

\begin{table}[h]
\begin{ruledtabular}
\centering
\begin{tabular}{cccccc}
Decay & $\overline{E}$ &    DTAS   &  JEFF     & ENDF          \\ 
 &  &    [keV]    &  [keV]     & [keV]   \\\hline

$^{96\text{gs}}$Y & 
\begin{tabular}{@{}c@{}}$\gamma$\\$\beta$\\$e^{-}$ \end{tabular} & 

\begin{tabular}{@{}c@{}}
\rule{0pt}{3ex}
66.8$^{+12.4}_{-1.5}$\\\rule{0pt}{3ex} 3193.0$^{+2.4}_{-18.6}$\\\rule{0pt}{3ex} 15.6$^{+24.9}_{-3.2}$
\end{tabular} & 

\begin{tabular}{@{}c@{}}
\rule{0pt}{3ex}
80.1(44) \\ \rule{0pt}{3ex} 3180.6(200) \\ \rule{0pt}{3ex} 22.1(19)
\end{tabular} & 

\begin{tabular}{@{}c@{}}
\rule{0pt}{3ex}
80.1(44)  \\ \rule{0pt}{3ex} 3184.0(173) \\ \rule{0pt}{3ex} 22.2(44)\footnote{\label{AUnote} This value also includes Auger electrons.}
\end{tabular} 

\begin{tabular}{@{}c@{}}
  \\  \\ 
\end{tabular} \\ \hline

$^{96\text{m}}$Y & 

\begin{tabular}{@{}c@{}}$\gamma$\\ $\beta$ \\ $e^{-}$ \end{tabular} & 

\begin{tabular}{@{}c@{}}
\rule{0pt}{3ex}
4669.2$^{+20.6}_{-12.1}$ \\ \rule{0pt}{3ex} 1720.5$^{+5.3}_{-8.5}$ \\ \rule{0pt}{3ex} 17.7$^{+1.2}_{-2.7}$
\end{tabular} & 

\begin{tabular}{@{}c@{}}
\rule{0pt}{3ex}
4479.1(823) \\ \rule{0pt}{3ex} 1821.2(1607)  \\\rule{0pt}{3ex} 29.7\footnote{\label{Errnote} No error value is given in the database.}
\end{tabular} & 

\begin{tabular}{@{}c@{}}
\rule{0pt}{3ex}
4308.4(3)  \\ \rule{0pt}{3ex} 1602.0(1625) \\ \rule{0pt}{3ex} 28.2(47)$^{\text{\ref{AUnote}}}$
\end{tabular}


\\ 

\end{tabular}
\caption{\label{Mean_exp} Average $\gamma$, $\beta$ and conversion electron energies of the decays of $^{96\text{gs,m}}$Y. The present TAGS results are compared with the values available in the ENDF/B-VII.1~\cite{ENDF} and JEFF-3.3~\cite{JEFF33} databases.}
\end{ruledtabular}
\end{table}

\subsection{$\beta$ spectra}\label{beta_spec}

\begin{figure}[h]
\begin{center}
\includegraphics[width=0.5 \textwidth]{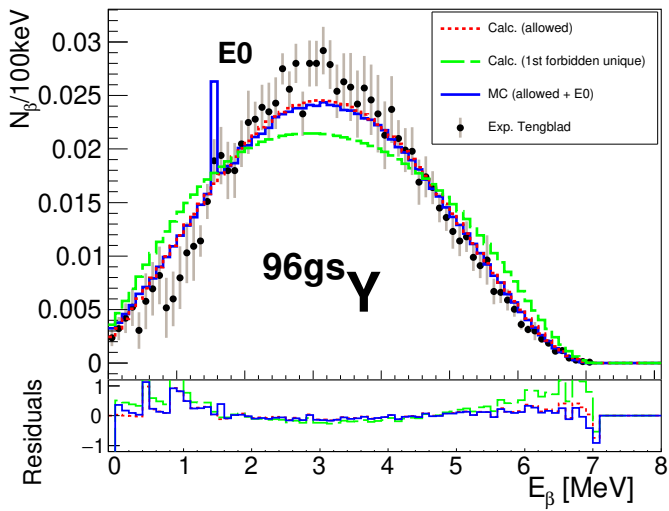} 

\caption{Electron spectrum for the decay of $^{96\text{gs}}$Y from a MC simulation with the $\beta$-intensity distributions obtained in this work (solid blue line) compared with the experimental data of Tengblad \textit{et al.} \cite{Olof_beta} (black points with error bars). The spike corresponds to the $E0$ transition from the level at 1581.6~keV. Note that we do not apply any experimental resolution to the MC $\beta$ spectra. Additional calculations without conversion electrons (dotted red) and assuming first forbidden unique $\beta$ shapes (dashed green) are also included for comparison. All spectra are normalized to 1. The relative deviations between calculated and experimental spectra are shown.}
\label{beta_spectra}
\end{center}
\end{figure}

Finally, we have employed the $\beta$-intensity distribution obtained in the present work for the decay of $^{96\text{gs}}$Y to determine the corresponding $\beta$-energy spectrum associated with this decay by means of subroutines from the \textit{logft} program of NNDC~\cite{logftNNDC}. In Fig.~\ref{beta_spectra} we compare the resulting spectrum with that measured by Tengblad \textit{et al.}~\cite{Olof_beta} mentioned earlier. Discrepancies are found in the range 2-4~MeV between the experimental $\beta$ spectrum and that calculated with our TAGS results, in line with the disagreement found in previous works for $^{86}$Br, $^{91}$Rb~\cite{Simon_PRC}, $^{87,88}$Br, $^{94}$Rb~\cite{vTAS_PRC}, $^{137}$I and $^{95}$Rb~\cite{PRC_BDN}. In our calculations we have assumed the shape of allowed transitions for all decay branches. Due to the dominance of the forbidden  0$^{-}\rightarrow$ 0$^{+}$ ground state to ground state transition in the decay of $^{96\text{gs}}$Y, we have also tested the assumption of first-forbidden unique transitions, observing an even larger disagreement with the experimental spectrum. Provided that conversion electrons were also measured together with $\beta$ electrons by Tengblad \textit{et al.}, in order to evaluate their influence in this comparison we have determined the electron spectrum simulated with the modified DECAYGEN event generator presented before. This allowed us to include the contribution of conversion electrons together with $\beta$ electrons (assuming again allowed $\beta$ shapes). As shown in Fig.~\ref{beta_spectra} the global shape of the spectrum does not change significantly, but a conversion electron peak is seen in the same position as in the experimental spectrum, corresponding to the 1581.6~keV level. Note that we do not apply any experimental resolution to the MC electron spectrum.

\section{Conclusions}\label{Conclusions}

In summary, we have studied the $\beta$ decays of $^{96\text{gs,m}}$Y by means of the TAGS technique. The isomeric 8$^{+}$ state was separated from the ground state with the JYFLTRAP double Penning trap system, which allowed us to study both decays separately, in contrast with recent studies where only one component or both decays mixed were investigated. The first excited state in the daughter nucleus $^{96}$Zr, a 0$^+$ that de-excites through conversion electrons in competition with pair production, was carefully treated in the response functions of our TAGS analyses. The conversion electron emission was found to have an impact on the efficiency of the $\beta$ detector, and the positron annihilation photons were clearly seen in our spectrometer, thus changing dramatically the shape of the response to this 0$^+$ level. The strong ground state to ground state feeding transition observed in the decay of $^{96\text{gs}}$Y was found to be sensitive to these effects, overlooked in recent independent TAGS studies. We have determined a ground state $\beta$ intensity of $96.6_{-2.1}^{+0.3}\%$, slightly larger than the previously reported value of 95.5(5)$\%$ but compatible within uncertainties. However, the impact of this change in reactor antineutrino summation calculations, where the decay of $^{96\text{gs}}$Y plays a major role between 5 and 7~MeV, was found to be small. A minor impact of the present results in reactor decay heat summation calculations was also observed. However, the uncertainties on the $\beta$ feeding of these two nuclei have been reduced by the new measurements presented here. This is reflected in the new average $\beta$ and $\gamma$ energies and their uncertainties, which are diminished significantly. The tools developed for the precise evaluation of the conversion electron branch will be applied to the study of the decay of the ground state of $^{98}$Nb, another case of interest for antineutrino spectrum studies with a strong $E0$ line.

The TAGS analysis of the decay of the 8$^+$ isomer confirmed the dominant population of the 8$^+$ state at 4389.8~keV excitation energy in $^{96}$Zr. However, previously unseen $\beta$ intensity was determined between 6 and 8~MeV, showing a clear Pandemonium effect in the high-resolution spectroscopy data available in ENSDF~\cite{NDS_96} and leading to average $\beta$ and $\gamma$ values differing by more than 100~keV and 200~keV respectively with respect to evaluated decay databases.

The segmentation of our spectrometer allowed us to investigate the application of the TAGS analysis methodology to the $M_m$-gated spectra. We have used a modified DECAYGEN event generator~\cite{TAS_decaygen} to construct the response function for each $M_m$ for the decay of $^{96\text{m}}$Y, and the results of the de-convolution of $M_m$=5,6 gated TAGS spectra were found to be in good agreement with the $\beta$-intensity distribution determined in the normal TAGS analysis. This method will be useful for future measurements of cases where the ground states and the isomeric states are very close in energy and have very different spin-parity values. 

As a final comment, we would like to stress that the conclusion recently derived in Ref.~\cite{PLB_96Y_2021} about the capabilities of modern high-resolution $\gamma$-spectroscopy HPGe arrays to overcome the Pandemonium effect, is totally case dependent, as can be seen from the results presented in this work. For a decay level scheme with low $\gamma$-multiplicity cascades and a relatively low number of levels involved, as in the case of $^{96\text{gs}}$Y, it is known that high-resolution $\gamma$-spectroscopy measurements do not necessarily suffer a dramatic bias. However, for cases with high $\gamma$-multiplicity cascades, as in the case of $^{96\text{m}}$Y, as well as in cases with large level densities and large $Q_{\beta}$ energy windows, where the $\beta$-intensity distribution is very fragmented, even the cutting-edge $\gamma$-spectroscopy arrays are handicapped, because of the limited efficiency of HPGe detectors and the characteristics of the technique, that relies on the detection of $\gamma$ rays in coincidence. Sufficiently far from stability, this approach always implies the loss of some $\gamma$ rays involved in the de-excitation cascades, thus shifting the deduced $\beta$-feedings as stated by Hardy \textit{et al.} in 1977~\cite{Pandemonium}.


\begin{acknowledgments}
This work has been supported by the CNRS challenge NEEDS and the associated NACRE project, the CNRS/IN2P3 PICS TAGS between Subatech and IFIC, and the CNRS/IN2P3 Master projects Jyv\"askyl\"a and OPALE. This work has also been supported by the Spanish Ministerio de Econom\'ia y Competitividad under Grants No. FPA2011-24553, No. AIC-A-2011-0696, No. FPA2014-52823-C2-1-P, No. FPA2015-65035-P, No. FPI/BES-2014-068222, No. FPA2017-83946-C2-1-P, and No. RTI2018-098868-B-I00 and the program Severo Ochoa (SEV-2014-0398); by the Spanish Ministerio de Educaci\'on under Grant No. FPU12/01527; by the Spanish Ministerio de Ciencia e Innovaci\'on under Grant No. PID2019-104714GB-C21; by the European Commission under CHANDA project funded under FP7-EURATOM-FISSION Grant No. 605203; the FP7/ENSAR Contract No. 262010; the SANDA project funded under H2020-EURATOM-1.1 Grant No. 847552. V.G. acknowledges the support of the Polish National Agency for Academic Exchange (NAWA) under Grant No. PPN/ULM/2019/1/00220 and of the National Science Center, Poland, under Contract No. 2019/35/D/ST2/02081. W.G. acknowledges the support of the U.K. Science and Technology Facilities Council grant No.ST/P005314/. This work was supported by the Academy of Finland under the Finnish Centre of Excellence Programme 2012-2017 (Project No. 213503, Nuclear and Accelerator-Based Physics Research at JYFL). Support from the IAEA Nuclear Data Section in acknowledged. Authors thank Tibor Kib\'edi for help with the new BrIcc version. 
\end{acknowledgments}

\bibliography{96Y}

\end{document}